\documentclass[twocolumn]{aastex631}

\newcommand{\sgn}{\mathop{\mathrm{sgn}}}
\usepackage{amsmath}

\usepackage[super]{nth}

\received{3 October 2024}
\revised{31 March 2025}
\accepted{16 April 2025}

\shorttitle{CHILES Con Pol Survey}
\shortauthors{CHILES Con Pol Collab}

\begin{document}

\title{The CHILES Continuum \& Polarization Survey-I: Survey Design \& Noise Characterization}

\correspondingauthor{Nick Luber}
\email{nicholas.m.luber@gmail.com}

\author{Nicholas M. Luber}
\affiliation{Department of Astronomy, Columbia University, Mail Code 5247, 538 West 120th Street, New York, NY 10027}

\author[0000-0001-7095-7543]{Min S. Yun}
\affiliation{Department of Astronomy, University of Massachusetts, Amherst, MA 01003, USA}

\author{Hansung B. Gim}
\affiliation{Department of Physics, Montana State University, P.O. Box 173840, Bozeman, MT 59717, USA}

\author{Daniel Krista-Kelsey}
\affiliation{Department of Astronomy, University of Massachusetts, Amherst, MA 01003, USA}

\author{D.J. Pisano}
\affiliation{Department of Astronomy, University of Cape Town, Private Bag X3,
Rondebosch 7701, South Africa}

\author{Emmanuel Momjian}
\affiliation{National Radio Astronomy Observatory, P.O. Box O, Socorro, NM 87801, USA}

\author{Chris Hales}
\affiliation{National Radio Astronomy Observatory, P.O. Box O, Socorro, NM 87801, USA}

\begin{abstract}

We introduce and describe the CHILES Continuum \& Polarization (CHILES Con Pol) Survey, a 1000 hour 1.4 GHz wideband full polarization radio continuum deepfield with the Very Large Array (VLA), commensurate with the CHILES HI deepfield. We describe the observational configuration, outline the calibration of the data, and discuss the effect of Radio Frequency Interference on different observing epochs. In addition, we present a novel radio continuum imaging strategy, using well known baseline subtraction techniques in radio spectral data, and discuss the applications to the removal of artifacts from sources far from the field center. Additionally, we discuss the nature of a low-level image-wide offset, the so-called ``negative bowl" and simulate our observations in order that we may both properly understand and correct for this artifact. Using these techniques, we present the first total intensity image of the data which achieves an r.m.s. noise of 1.3 $\mu$Jy beam$^{-1}$ with a synthesized beam of 4.5\arcsec x 4.0\arcsec, the most sensitive L-band image ever taken at this resolution. We then place this image into the broader context of 1.4 GHz radio continuum surveys in the literature, in terms of image sensitivity and fidelity, and $\mu$Jy level source counts and P(D) analysis.

\end{abstract}

\keywords{radio astronomy -- imaging techniques -- extragalactic astronomy}

\section{Introduction}\label{sec:intro}

\quad The observation of radio continuum emission can provide crucial measurements for understanding the evolution of galaxies across cosmic time. Radio emission can be used to estimate star-formation rates by means of the FIR-radio correlation \citep{condon92,yun01}. The use of this relationship is motivated by the scenario in which UV radiation from young bright OB stars is absorbed by dust that produce FIR emission while supernovae from these same stars produces non-thermal relatavistic electrons moving within a galaxy's magnetic field producing radio synchrotron emission \citep{bell03}. Radio emission can also arise from active galactic nuclei (AGN) forming radio jets and large radio lobes. These lobes can interact with the interstellar medium within the host galaxy resulting in negative or positive feedback \citep{cresci18}, and the medium surrounding the galaxy providing insight into the density and magnetism of intra-galactic matter \citep{fanaroff74, hardcastle20, muller21}. Additionally, the measurement of polarized radio emission can be used to characterize magnetic fields, in environments spanning from the jets of AGN to cosmological scales \citep{laing80, kolatt98}.

\quad These physical phenomenon have been studied with several different large-field 1.4 GHz surveys of varying sensitivity and resolution, e.g., in the northern hemisphere, the NRAO VLA Sky Survey \citep[NVSS,][]{condon92} which achieved sensitivity of approximately 0.5 mJy beam$^{-1}$ with 45$\arcsec$ resolution and the Faint Images of the Radio Sky at Twenty Centimeters Survey \citep[FIRST,][]{becker95} which achieved a sensitivity of approximately 0.15 mJy beam$^{-1}$ with a resolution of 5$\arcsec$, and in the southern hemisphere, the Sydney University Molonglo Sky Survey \citep[SUMSS,][]{bock99} which achieved a sensitivity of approximately 1 mJy beam$^{-1}$ with a resolution of 45$\arcsec$, at the lower frequency of 0.84 GHz. Currently, the VLA is undertaking the Very Large Array Sky Survey \citep[VLASS,][]{lacy20} which is expected to achieve 70.4 $\mu$Jy beam$^{-1}$ with 2.5$\arcsec$ resolution, over a bandwidth spanning 2 - 4 GHz. These large-field surveys have been able to detect many thousands of different continuum sources at differing levels of intensity, resolutions, and spatial extents. These large-field surveys are ideally complemented by a radio continuum survey that has both excellent resolution ($<$5$\arcsec$) and sensitivity ($<$2 $\mu$Jy beam$^{-1}$), which would be able to probe the radio continuum physics of individual sources out to higher redshifts, as well as low-level diffuse structures from low redshift sources that could be missed by less sensitive all-sky surveys.

\quad To achieve this, such a survey requires a significantly longer integration per field and thus would likely be limited to a single field, or a few fields. It is also critical that there exist significant multi-wavelength coverage to properly characterize the Spectral Energy Distribution (SED), and subsequently properly characterize the nature of the radio emission. With these in mind, it is the natural decision to conduct a deep radio continuum survey that overlaps with the COSMOS field, a 1.4$^{\circ}$ x 1.4$^{\circ}$ equatorial field that has extensive multi-wavelength coverage \citep{scoville07}. The COSMOS field has observations that include low-frequency radio coverage \citep{smolcic14,schinnerer07,smolcic17}, the sub-millimeter and infrared \citep{lefevre20,oliver12}, the near-infrared and optical spectroscopy \citep{capak07,lefevre15,davies15} and imaging \citep{grogin11,mccracken12}, and the ultraviolet and X-ray \citep{zamojski07,civano16,marhcesi16}. Additionally, using this extensive multi-wavelength coverage the COSMOS team has catalogues with hybrid measurements such as the color defined by the near-ultraviolet and R-band magnitudes (NUV-R), and derived properties such as redshifts, both spectroscopic and photometric, star-formation rates, and stellar masses \citep{davies15,laigle16,weaver22}.

\quad The COSMOS field has been previously observed at 1.4 GHz regime by \citet{schinnerer07} with the VLA in its most extended (A) and second most compact (C) configurations. They achieved an r.m.s. noise of approximately 10 $\mu$Jy per 1.5$\arcsec$ beam. The project was a mosaic across the entire COSMOS region with a total of 240 observing hours. More recently, the COSMOS field was observed by the MeerKAT International GHz Tiered Extragalactic Exploration Survey \citep[MIGHTEE,][]{jarvis16} using the new South African MeerKAT telescope. The MIGHTEE survey is a multi-pointing observation that encompasses most of the COSMOS field, as well as mosaics over the XMM-Newton Large Scale Structure, E-CDFS, and ELAIS-S1 fields. They produce images that have a factor $\approx$5 worse resolution than in \citet{schinnerer07}, but a factor $\approx$5 better sensitivity \citep{heywood22}. Additionally, the COSMOS field has been observed using very long baseline interferometry, specifically with the Very Long Baseline Array, which achieves milli-arcsecond resolution and allows for in-depth studies into AGN fraction of radio source counts \citep{herrera17,herrera18}.

\quad In this work, we introduce the CHILES Continuum Polarization (CHILES Con Pol) Survey, a full polarization 1.4 GHz radio continuum survey taken by the VLA in its second most extended (B) configuration with 1027 hours of total observing time. The CHILES Con Pol survey is taken simultaneously with the CHILES survey and is a single pointing ($\approx$0.25 square degrees) coincident with a section of the COSMOS field. The observations were expected to, and successfully yield images with an r.m.s. noise of $\approx$ 1-2 $\mu$Jy per beam, and resolution of 4.5$\arcsec$, making them the most sensitive observations at this resolution in L-band (1-2 GHz). In Section \ref{sec:observations}, we discuss the survey design and instrumental set-up, and in Section \ref{sec:processing}, we describe the calibration, imaging, and Radio Frequency Interference environment of the observations. In Section \ref{sec:results}, we show the total intensity image of the field and discuss the applicability of the data to different scientific pursuits and in Section \ref{sec:disc}, we discuss our results in the context of other radio continuum surveys of COSMOS. In this work we assume a standard flat $\Lambda$CDM cosmology with H$_{0}$ = 70, $\Omega_{\Lambda}$ = 0.7 and $\Omega_{M}$ = 0.3.

\begin{deluxetable}{cc}
\tablecaption{Summary of Observations\label{tab:ccp_prop}}
\tablecolumns{1}
\tablenum{1}
\tablewidth{0pt}
\tablehead
{
\colhead{a} &
\colhead{b}
}
\startdata
Target Pointing & 10h01m24s +02$^{\circ}$21$\arcmin$00$\arcsec$\tablenotemark{a} \\
Observation Dates & 2013/10/25 - 2019/04/11\tablenotemark{b} \\
Total Integration Time & 1027 hrs\tablenotemark{c,d} \\
Bandpass Calibrator & 3C286 \\
Complex Gain Calibrator &  J0943–0819 \\
Dump Time & 8s \\
Spectral Window Setup & 4 x 128MHz\tablenotemark{e} \\
Channel Width & 2 MHz \\
\enddata
\tablenotetext{a}{In the J2000 coordinate standard.}
\tablenotetext{b}{The observations were taken over five different epochs, corresponding to five consecutive VLA-B array configurations. This results in the observations being taken over different parts of the year and having a mix of both day and night time observations.}
\tablenotetext{c}{The total integration time is split amongst the five observing epochs with 188, 220, 190, 237, 192 hours, respectively, in a total of 210 scheduling blocks.}
\tablenotetext{d}{Individual sessions have total observation times ranging from 1 - 8 hours, with a mean time of 4.75 hours.}
\tablenotetext{e}{The frequencies for the first channel in the four spectral windows are 1000, 1384, 1640, and 1768 MHz, respectively.}
\end{deluxetable}

\section{Survey Design}\label{sec:observations}

\begin{figure}
\begin{center}
\includegraphics[width=\columnwidth]{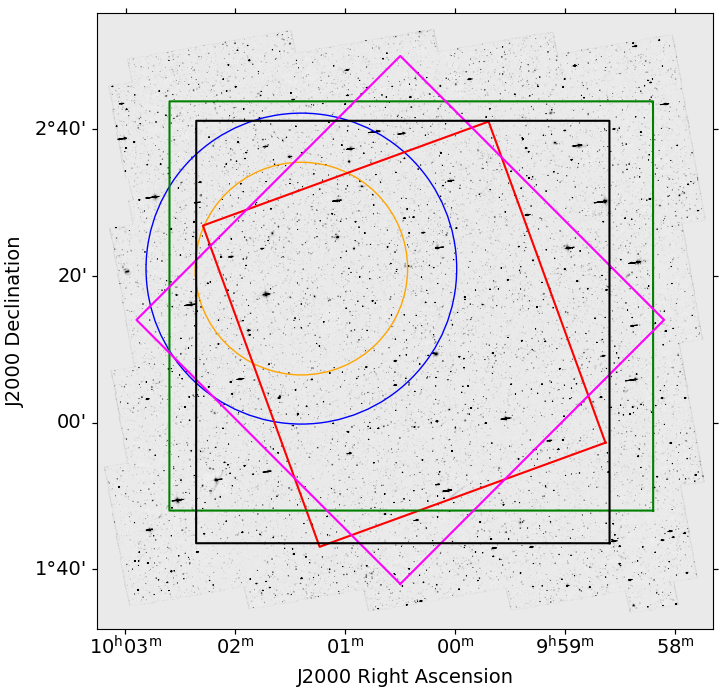}
\caption{The CHILES Con Pol field, defined as the 20\% (blue circle) and 50\% (orange circle) of the VLA primary beam at 1.4 GHz, overlaid on the COSMOS HST coverage \citep{koekemoer07}. In addition, to provide context for the extensive multi-wavelength coverage in the COSMOS field, we present the footprints of just a few of the many surveys the JWST COSMOS-Web fied-of-view is presented in red \citep{casey2022}, the AxTEC sub-millimeter survey is presented in magenta \citep{aretxaga11}, the CFHT NUV survey is presented in black \citep{capak07}, and the GAMA G10 region of reprocessed COSMOS photometry and spectroscopic is shown in green\citep{andrews17}.}
\label{fig:CHILES_field}
\end{center}
\end{figure}

\quad The CHILES Continuum Polarization (CHILES Con Pol) Survey is a commensurate survey to the COSMOS HI Large Extragalactic Survey (CHILES), which aims to study neutral hydrogen emission out to redshifts $z$ $\approx$ 0.45 \citep{fernandez13, fernandez16}, and the CHILES Variable and Explosive Radio Dynamic Evolution (CHILES VERDES), a time domain project making use of the CHILES Con Pol data products to explore variations on the timescales of days at an unprecedented sensitivity \citep{sarbadhicary21}. The CHILES pointing was chosen to be coincident with the COSMOS field, thus able to utilize the extensive multi-wavelength coverage of the COSMOS survey \citep{scoville07}. The pointing is centered on J2000 10h01m24s +02d21m00s, and in Figure~\ref{fig:CHILES_field} we show the CHILES Con Pol field of view overlaid on an HST map along with the footprint of several of the deep multi-wavelength surveys that have been undergone in the COSMOS field. The CHILES field is off center from the COSMOS field to avoid strong continuum sources, our strongest in-field continuum source has a flux density of 15 mJy, to reduce the impact of any residual calibration and imaging artifacts.

\quad The CHILES surveys were taken by the Karl G. Jansky Very Large Array (VLA) operated by the National Radio Astronomical Observatory\footnote{The National Radio Astronomy Observatory is a facility of the National Science Foundation operated under cooperative agreement by Associated Universities, Inc.} (NRAO). In 2011, the VLA was equipped with new wide-bandwidth receivers, including a new updated L-band (1-2 GHz) receiver with modern electronics, as part of the Expanded VLA (EVLA) project \citep{evla}. Additionally, as a part of the upgrade, was the new correlator, the Wideband Interferometric Digital ARchitecture (WIDAR) which now allows for 64 independent subbands to be simultaneously observed, and subsequently processed \citep{carlson00}.

\quad The spectral line CHILES observations require 60 of the 64 available baseline board pairs to deliver fifteen 32 MHz subbands with a total of 30,720 channels across 480 MHz of bandwidth with dual polarization (RR, LL) products. The remaining four baseline board pairs were configured to deliver four 128 MHz wide subbands, each with 64 channels and full polarization products (RR, LL, RL, LR). Therefore, the resulting total bandwidth of the CHILES Con Pol data is 512 MHz. These four spectral windows are spread across L-band with the central channel in each spectral window being 1064, 1448, 1704, and 1832 MHz, respectively, and are chosen to properly sample L-band frequencies while also avoiding regions that are known to be heavily impacted by Radio Frequency Interference (RFI). The correlator integration time used for the observations is eight seconds to ensure that time smearing is minimal, while also minimizing the total data volume.

\begin{figure*}
\begin{center}
\includegraphics[width=\textwidth]{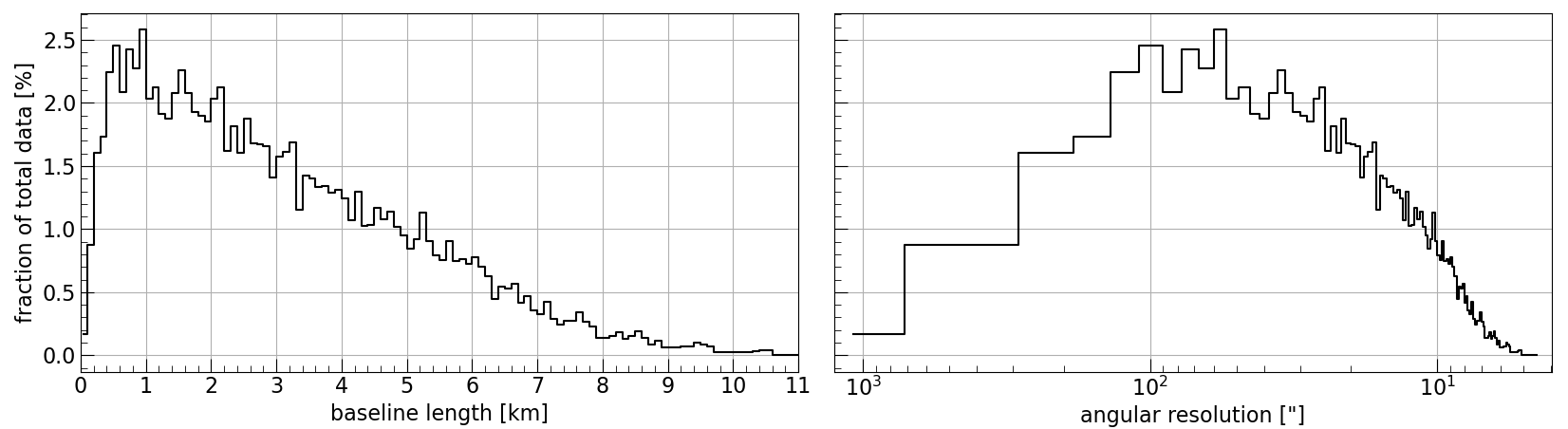}
\caption{\textbf{\textit{Left:}} The percentages of total visibilities as a function of baseline length, with the data put into bins of width 100m. \textbf{\textit{Right:}} The percentages of total visibilities, using the same bins as the plot to the left, as a function of the angular scale to which they are sensitive to emission.}
\label{fig:base_dist}
\end{center}
\end{figure*}

\quad The CHILES Con Pol observations were taken with the VLA in its second most extended array form, the B-configuration. The VLA B-configuration has a maximum baseline length of $\approx$11km, corresponding to an angular resolution of 4.8$\arcsec$ at 1.4 GHz. In Figure~\ref{fig:base_dist}, we show the percentage of the total CHILES Con Pol observations as a function of baseline length (left) and the percentages with the corresponding angular sky resolution (right). These percentages were taken from the idealized case that no data were flagged due to RFI, and that all 27 antennas were functional for every observation. This is an approximation as it is known that RFI will preferentially affect some baselines, and that some antennas are out of the array during different observing sessions. However, on average, Figure~\ref{fig:base_dist} is an accurate representation of the distribution of baselines for the full survey. The VLA B-configuration was chosen as it it has the excellent resolution and can resolve the HI kinematics in higher redshift sources, while also being less affected by RFI than the more compact VLA configurations. Additionally, for the CHILES Con Pol survey that will approach unprecedented sensitivity, if our resolution were any worse, the data would have been fully confusion limited. As a result, the B-configuration allows for increased source detection and characterization for the CHILES Con Pol survey.

\begin{figure*}
\begin{center}
\includegraphics[width=\textwidth]{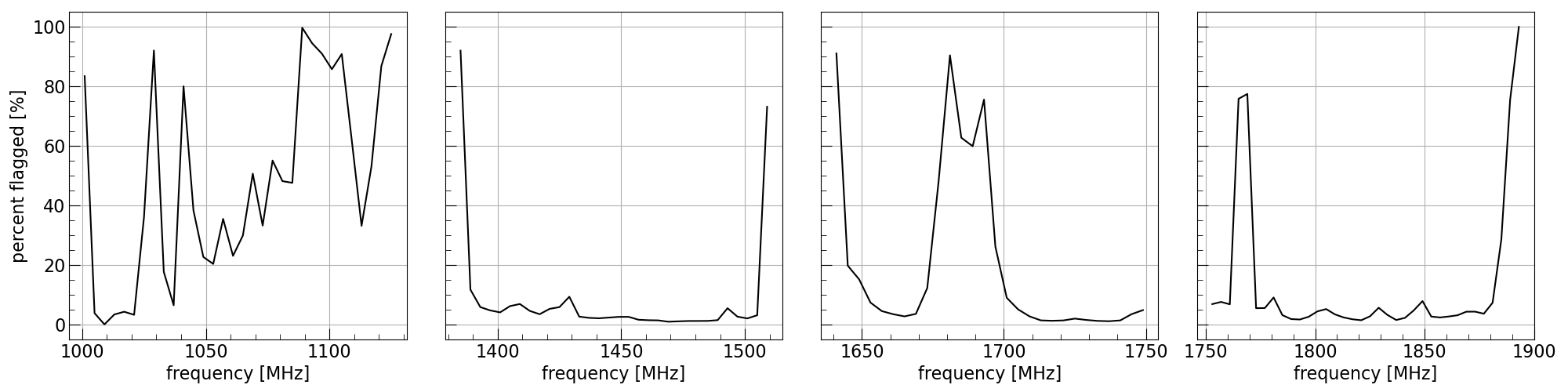}
\caption{The percent of the total visibilities flagged for the entire CHILES Con Pol database. The four panels of the plot represent the split for the four spectral windows used in the observations.}
\label{fig:1000hr_flagstat}
\end{center}
\end{figure*}

\begin{figure*}
\begin{center}
\includegraphics[width=\textwidth]{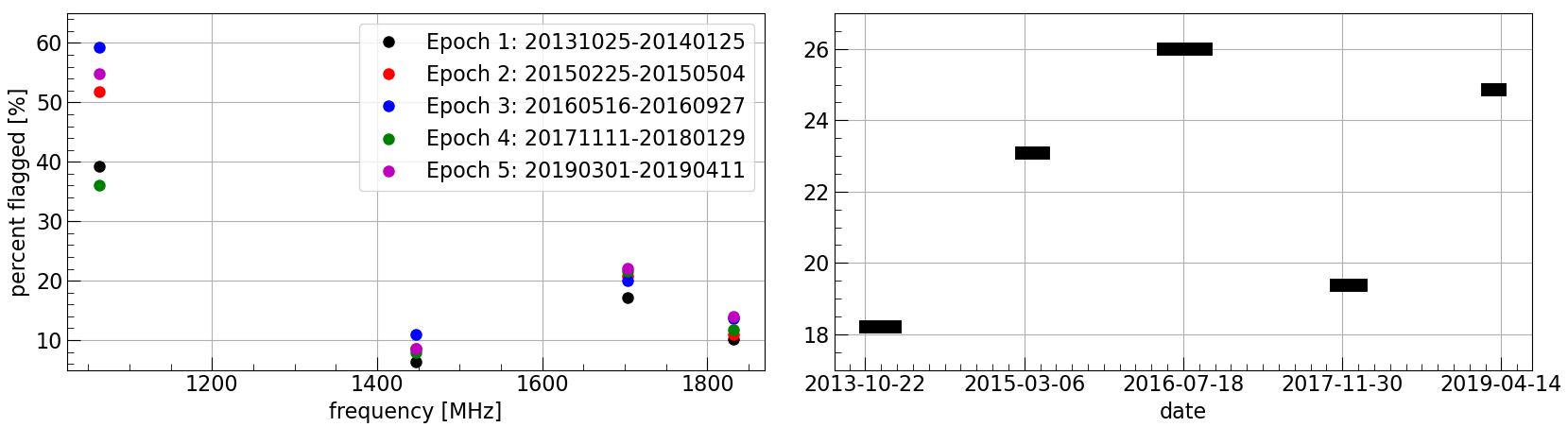}
\caption{\textbf{\textit{Left:}} The percentages of total visibilities flagged per each 128 MHz spectral window, and per the consecutive VLA-B configurations.\textbf{\textit{Right:}} The percent of total visibilities flagged per the consecutive VLA-B configurations.}
\label{fig:PerEpoch_flagstat}
\end{center}
\end{figure*}

\section{Data Processing}\label{sec:processing}

\subsection{Calibration}\label{sec:cal}

\quad The calibration of the data is done entirely using the Common Astronomical Software Application \citep[CASA,][]{casa}, version 4.7, and employs standard procedures and diagnostics. Before we begin the bandpass, complex gain, and polarization calibration, we perform perform a number of processing steps that provide an overview of the data and proper calibration formatting. These are summarized and undertaken in the following steps:

\begin{enumerate}
    \item We create a per session working directory, download the VLA Observers Log, and the data in its SDM-BDF format. The data is then imported into the standard CASA measurement set format while applying the online flags. The online flags and weather information are then plotted, and initial flagging statistics saved. This first measurement set includes the fifteen 32 MHz subbands of the CHILES data and the four 128 MHz CHILES Con Pol data. We then split off the four CHILES Con Pol 128 MHz subbands into their own measurement set and delete the full measurement set, as well as the SDM-BDF directory.
    \item We retrieve the specifics of the observations from the measurement set and write out the channel number-frequency relationship for each spectral window. We calculate the antenna position corrections, and plot the antenna positions and elevation versus time. We apply the pure zero, shadow, and auto-correlation flags, and hanning smooth the data. We then flag two channels at the beginning and end of each spectral window, to mitigate the effects from the edge of the bandpass, and flag the first scan after each field change, to ensure the antennae are on source for the entirety of the scans used for calibration.
    \item We create some intial plots for data inspection, specifically the amplitude versus frequency for the two calibrators and target field. We set the model for 3C286 using the L-band standard from \citet{perley13} and the position angle of 3C286 to 33$^{\circ}$ across the entire bandwidth. We then create weblogs for the flagging, and the data overview plots made subsequent to this step. Lastly, we create text files for manual flags and for the reference antenna, spectral window, and channel set-up to be used in subsequent steps.
\end{enumerate}

\quad After these initial steps, we perform the gain calibrations. These calibrations make use of several tasks custom made for the reduction of this data, specifically, \citep[\textit{interpgain},][]{interpgain} and \citep[\textit{antintflag},][]{antintflag}. Additionally, the gain calibration routine uses the custom task \textit{pieflag} \citep{pieflag}, which we will describe in more detail in Section \ref{sec:rfi}, in addition to the final flagging statistics for the target data. The gain calibration steps are summarized and undertaken in the following steps:

\begin{enumerate}
    \item Import the ionospheric corrections and generate the corresponding calibration table. Create three dimensional plots of time and frequency as the x and y axis, and amplitude as the color scale for all targets and spectral windows for the RL correlation. Using these plots, select the reference channels, and document and apply any channel or baseline flagging for all fields, and any time flagging for 3C286 and the reference channels, and report the flagging statistics after this step.
    
    \item Perform the initial phase calibration for 3C286, three channels wide around the reference channel, plot the primary calibrator initial phase solutions, and refine flagging, if required. Perform and plot the delay calibration, and refine flagging, if needed. Do the bandpass calibration, plot the solutions, and perform any neccessary additional flagging. We then do the final temporal gain calibration for 3C286, plot the solutions, and flag as required.
    
    \item We do an initial amplitude and phase calibration for J0943-0819, again three channels wide around the reference channel, plot the initial solutions, perform any neccessary flagging, and interpolate the gain when neccessary to account for any flagging. Next, we apply the antenna position corrections, bandpass, delay, and preliminary gain calibration to J0943-0819 and perform additional flagging using \textit{pieflag}.
    
    \item We now do the final gain calibration for J0943-0819. We then plot the final solutions and perform any additional flagging, if need be, and finally, interpolate the gain solutions on account of any of the additional flagging.
    
    \item We perform and plot the per-spectral window cross-hand delay calibration solutions, and do any additional flagging. Next, assuming zero source polarization, we perform, and plot the solutions of, the leakage calibration, and perform any additional flagging. We then do the RL phase calibration, and subsequently plot the solutions and flag as necessary.
    
    \item We now scale the amplitude gains for J0943-0819 and plot the rescaled amplitude data. We then apply the full calibration to the two calibrators, and plot and inspect the data.  If the calibration is satisfactory, we apply the calibration to the target field, and use \textit{pieflag} on the calibrated target data. Lastly, we create plots of amplitude vs. frequency and UV distance and report the final flagging statistics.
    
\end{enumerate}

\subsection{Radio Frequency Interference}\label{sec:rfi}

\quad The data were flagged on a per-session basis using the flagging agent, \textit{pieflag}, which is run in the CASA environment \citep{pieflag}. \textit{pieflag} runs by comparing several different statistics on visibility amplitudes for each channel, to a user-determined RFI-free channel, on a baseline by baseline basis. For a full description of the algorithm see \citet{middleberg06}. This results in 21.6\% of the total target data being flagged. In Figure~\ref{fig:1000hr_flagstat}, we show the percent of total visibilities flagged per channel, where each of the panels corresponds to a spectral window. The lowest frequency spectral window is clearly the most affected by RFI due to aircraft navigation radars. Similarly, the third spectral window is impacted by RFI from weather satellites. 

\quad The CHILES Con Pol observations were taken in a variety of times of the day, and season of the year. As a result, the different observing epochs have differing exposure to RFI sources and subsequent flagging statistics. The first and fourth observing epochs were observed, generally, in the early morning near sunrise in late fall/early winter, second and fifth observing epochs were observed in the middle of the night in late winter/early spring, and the third epoch was observed in the late afternoon/early evening in the summer time. In Figure~\ref{fig:PerEpoch_flagstat}, we can see that the flagging statistics for the observing epochs that were observed in similar seasons/times are remarkably similar, as to be expected. Interestingly, the best RFI conditions seem to be the early winter and early morning observations. The fact that the third observing epoch has the most data flagged is to be expected, as it was the most susceptible to daytime RFI, and the later half of the observing epoch has additional flags as a result of the field becoming closer to the sun, and shorter baselines being highly susceptible to solar RFI. Lastly, each spectral window varies similarly in time, indicating, that there is no preferential season/time for observations across the entire L-band, but rather, thee effects of worsening RFI are felt across all low frequencies.

\begin{figure*}
\begin{center}
\includegraphics[width=0.9\textwidth]{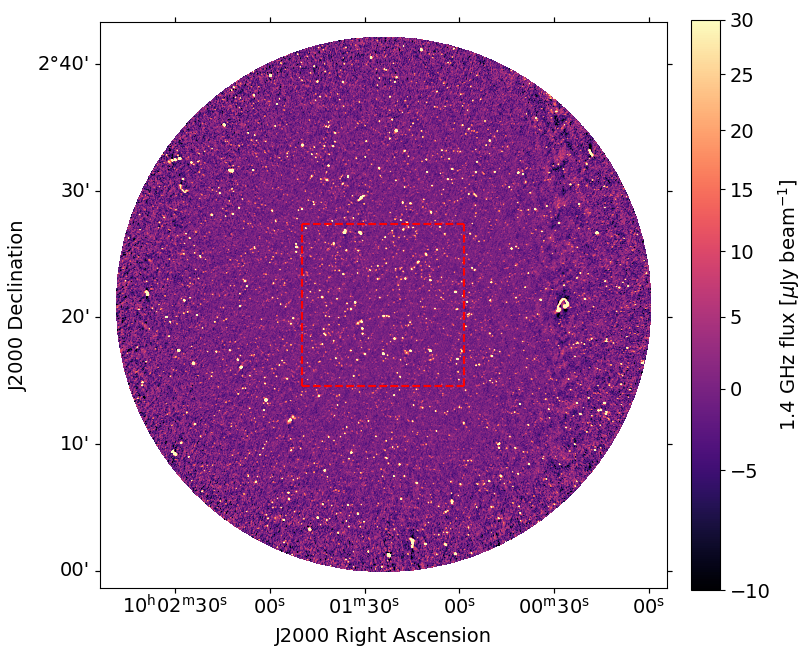}
\caption{The primary beam corrected CHILES Continuum Polarization total intensity map at the resolution of 4.5$\arcsec$ x 4.0$\arcsec$, out to the 20\% point of the VLA at 1.4 GHz. The central red dashed square, of side length 8.6$\arcmin$, corresponds to the region shown in the following Figure~\ref{fig:ccp_imap_zoom}.}
\label{fig:ccp_imap}
\end{center}
\end{figure*}

\begin{figure*}
\begin{center}
\includegraphics[width=0.9\textwidth]{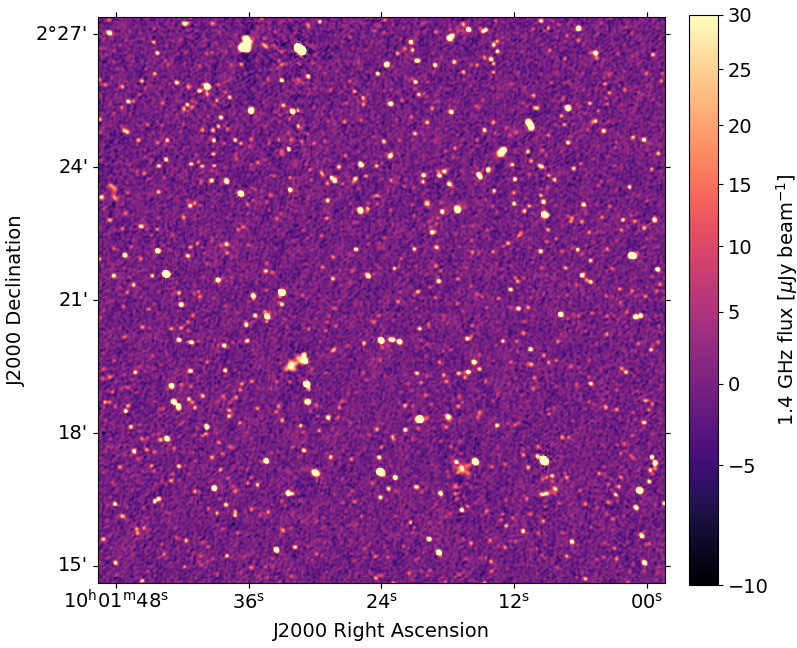}
\caption{The primary beam corrected CHILES Continuum Polarization total intensity map at the resolution of 4.5$\arcsec$ x 4.0$\arcsec$, for the red dashed square, of side length 8.6$\arcmin$, indicated in the previous Figure~\ref{fig:ccp_imap}.}
\label{fig:ccp_imap_zoom}
\end{center}
\end{figure*}

\subsection{Imaging} \label{sec:img}

\quad The production of accurate high dynamic range radio continuum total intensity images is a well studied problem. As surveys achieve higher and higher sensitivity, the largest source of errors in the images come from direction-dependent effects caused by time, frequency and direction-dependent effects in the telescope primary beam pattern, as well as antenna pointing errors \citep{heywood22}. The effects of these errors appear as deconvolutional errors where a source is unable to be CLEANed properly, and the total intensity image has significant artifacts from the PSF response to these sources.  These errors become more pronounced the farther away from field center a source is, as these are the places in which the telescope primary beam response varies the most. Additionally, the intrinsically stronger a source is, the stronger the residual sidelobes and subsequent artifacts.

\quad Several different methods exist to try and counter this effect. One such example is that of peeling, where the traditional approach is to subtract all sources from the field, except for the problematic source, phase rotate the data to be centered on this source, derive a direction-dependent calibration, apply the calibration and subtract the source from the data, and then return the data to its initial pointing and calibration with all other sources added back to the field \citep{williams19}. However, a technique such as this requires that sources be of a significant enough flux density to solve for the gain solutions. The CHILES Con Pol problematic sources have flux densities that do not meet this criterion for a single observing session, which is the scale on which this must be done, but are still strong enough to produce image artifacts in the final combined image.

\quad In order for CHILES Con Pol to produce accurate total intensity image reconstructions, we pursue a novel approach for dealing with these problematic sources in radio continuum data. The approach we introduce below is done entirely with CASA version 5.6, and was done using the West Virginia University High Performance Computing Cluster, Thorny Flat\footnote{Computational resources were provided by the WVU Research Computing Thorny Flat HPC cluster, which is funded in part by NSF OAC-1726534.}, and takes approximately one week of computation using 96 GB of memory.

\quad For each individual session, we create a square image with a side length of 1.7$^{\circ}$ with a Briggs robust value of $R= +$0.5, 1024 wplanes, four Taylor terms, and CLEAN down to the 1$\sigma$ level in the areas of known continuum sources. These sources are then subtracted from the field. We then phase shift to a source far from the field center (outside of our 1.7$^{\circ}$ image) and perform a uv-plane linear fit, per each spectral window, and subtraction of the problematic source. We repeat this process for two other problematic sources far from the field center. The use of a uv-plane linear fit is preferred as it is known to be capable of subtracting sources subject to imperfect bandpass calibration, such as our distant sources suffering from direction-dependent effects. As a result, subtracting them from the field produces higher quality images of the science field in a manner similar to peeling, without requiring the high signal-to-noise required by peeling to produce appropriate gain solutions. After this is done, we return the field to the proper phase center and restore the CLEANed sources.

\begin{figure}
\begin{center}
\includegraphics[width=\columnwidth]{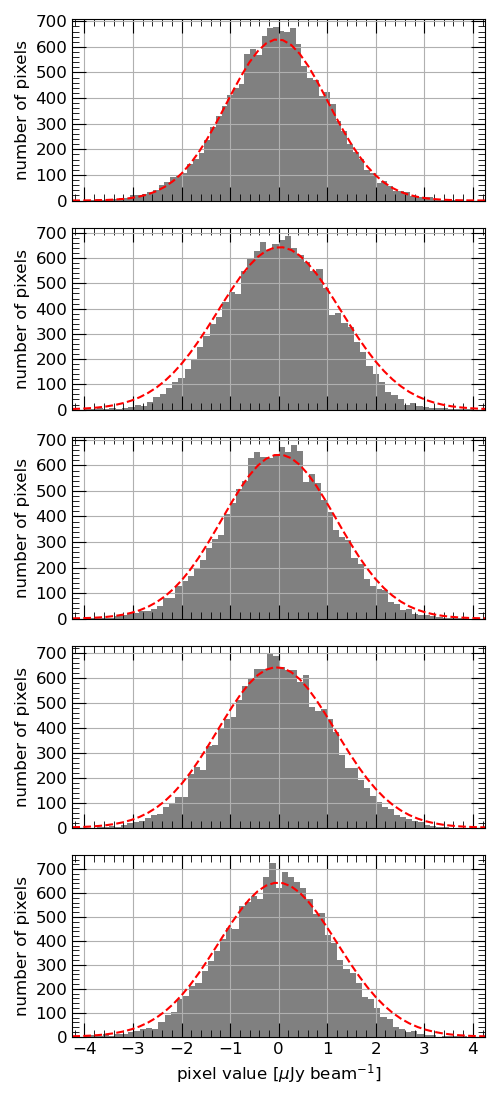}
\caption{The distribution of pixel values for the total intensity image over five blank sky regions 39, 44, 48, 49, and 57 arcminutes away from the phase center (shown here in descending plots) with sizes of 3\arcmin$\times$3\arcmin. The distributions are shown for the range $\pm$4 $\mu$Jy beam$^{-1}$ range with the bin size dictated by the Freedman-Diaconis rule. The red-dashed lines indicate the Gaussian fits to the distributions used in estimating the rms noise.}
\label{fig:BlankSkyRMS}
\end{center}
\end{figure}

\quad Following the per-session problematic source subtraction, we create a square image per observing epoch (see Table \ref{tab:ccp_prop} for the breakdown of sessions per epoch) with a side length of 1.7$^{\circ}$ with a Briggs robust value of $R=-$1, 1024 wplanes, four Taylor terms, and CLEAN down to the 1$\sigma$ level in the areas of known continuum sources. We then combine these five per-epoch images in the image plane using an inverse variance weighted mean.

\begin{figure*}
\begin{center}
\includegraphics[width=\textwidth]{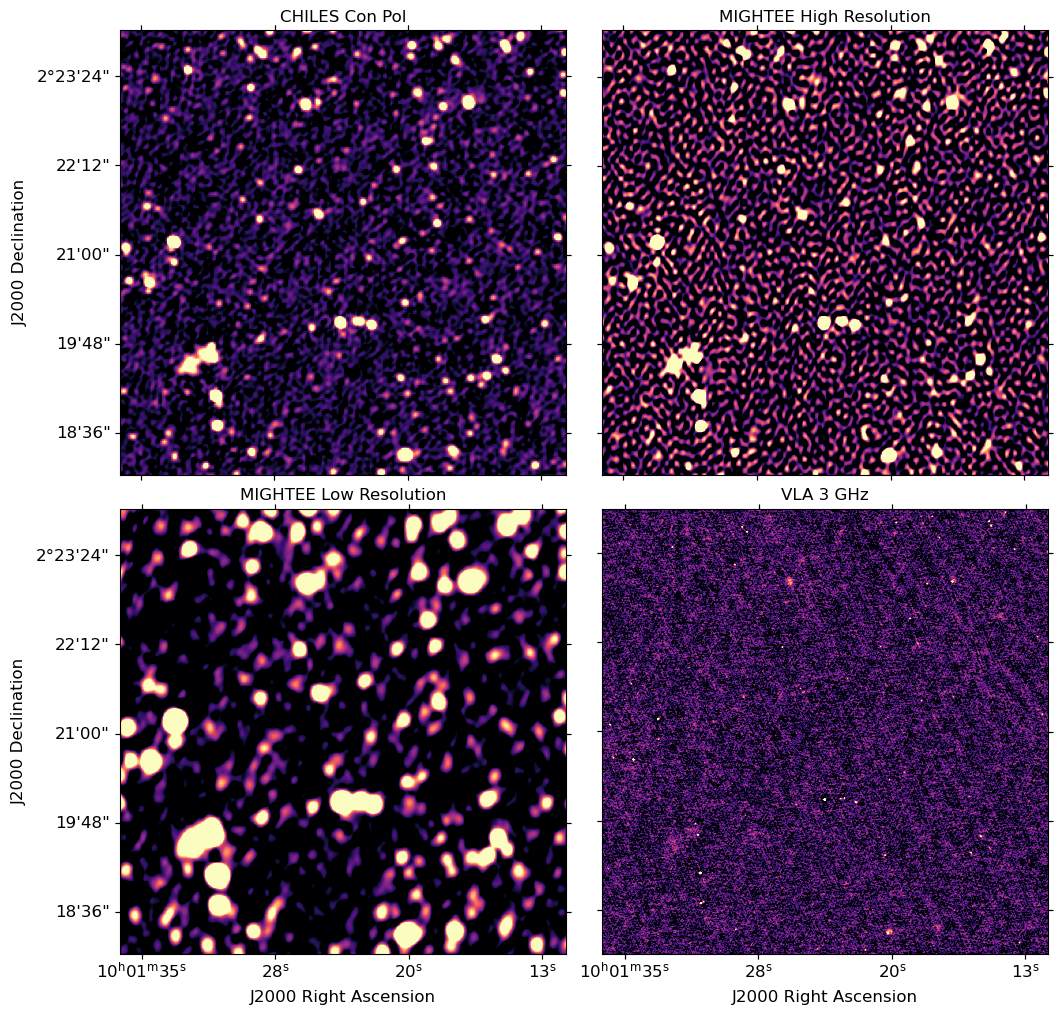}
\caption{The total intensity radio continuum images of a $3\arcmin \times 3\arcmin$ region for a detailed comparison among the CHILES Con Pol (top left), MIGHTEE high ($\theta=5\arcsec$; upper right) \& low resolution \citep[$\theta=8\arcsec.6$; lower left][]{heywood22}, and VLA-COSMOS 3GHz \citep[$\theta=0.75\arcsec$; lower right][]{smolcic17} data. The intensity ranges shown are between $-1$ and $+20$ $\mu$Jy for the first three panels and $-1$ and $+10$ $\mu$Jy for the VLA-COSMOS 3~GHz image.  The VLA-COSMOS 1.4~GHz survey data \citep{schinnerer07} is too shallow ($\sigma\approx10\,\mu$Jy) to offer a meaningful comparison and is not shown.}
\label{fig:ccp_fieldcompare}
\end{center}
\end{figure*}

\vspace{1cm}

\section{Results} \label{sec:results}

\quad In Figures \ref{fig:ccp_imap} and \ref{fig:ccp_imap_zoom}  we present the total intensity image using the CHILES Con Pol data, made from the full 1000 hours of total observing time.\footnote{These data are publicly available and can be found: 10.5281/zenodo.13877267} Figure~\ref{fig:ccp_imap} shows the full field of view of the observations (the 20\% point of the VLA primary beam response at 1.4 GHz), and a zoom in on the inner quarter of field of view is shown in Figure~\ref{fig:ccp_imap_zoom}. In this field, we detect approximately 1636 sources that will be characterized and matched with COSMOS multi-wavelength data in Gim et al. (in press). A visual inspection of the total intensity map without applying the primary beam cutoff makes it obvious that there is significant extra radio continuum emission within the VLA primary beam area, presumably arising from the high density of $\mu$Jy radio sources (see Section~\ref{sec:microJy} for further discussions) indicating that these ultra-deep radio continuum data are confusion-limited. 

\quad We estimate the intrinsic map noise in five featureless regions far outside the primary beam. The five featureless regions were at 39, 44, 48, 49, and 57 arcminutes away from the phase center and have sizes of 3\arcmin$\times$3\arcmin. For the total intensity image, the distributions for the five regions is shown in Figure \ref{fig:BlankSkyRMS}. We drew the histogram of flux densities after accumulating all signals in five regions to remove the stochastic effect. The bin width is robustly measured by applying the Freedman-Diaconis rule. Then, we applied the Gaussian fit to the histogram using the general purpose optimization function $optim$ in R to estimate the RMS noise, which yielded the RMS of 1.09, 0.79, 0.76, and 0.80 $\mu$Jy beam$^{-1}$ for I, Q, U, and V-stoke, respectively. The rms noise measured in the center of the image is significantly higher, 2.03 $\mu$Jy beam$^{-1}$ (see Table~\ref{tab:noise}), and we attribute this to additional ``noise" to the source confusion and imaging artifacts (see Section \ref{sec:extended}). We estimate the expected theoretical thermal noise by calculating the average System Equivalent Flux Density (SEFD) of the observations by using the known frequency-dependent SEFD values from \citet{memo204}, and the frequency-dependent number of visibilities as the weighting. Using this method, the theoretical thermal noise expected from the total number of visibilities is 0.63 $\mu$Jy beam$^{-1}$, which is much closer to the measured image noise far outside the primary beam, especially in the Q, U, and V-Stokes images where both confusion and deconvolution noises are expected to be much lower.

\quad The image presented in Figures \ref{fig:ccp_imap} and \ref{fig:ccp_imap_zoom} is of unprecedented sensitivity and resolution, but is affected by several sources of systematic errors. The most noticeable artifact in the image plane is a deconvolution failure around the extended source at 10h 25m and +2$^{\circ}$ 22', colloquially referred to as the ``Earmuffs" galaxy. This complex source is at the approximate 50\% point of the VLA primary beam resulting in an imperfect calibration. Additionally, there are some low-level spoke-like features emanating across the image (from SE to NW) that are a result of sources beyond the first null of the primary beam. Although these image-plane errors are noticeable, we still achieve close to Gaussian noise and within a factor of 4 of the theoretical expectation. Future data releases will aim to address these issues using more advanced imaging and deconvolution techniques, while also producing high fidelity images of other three Stokes planes.

\quad Any radio interferometric map, at a given resolution, with suitably long integration will reach a level where the predicted thermal noise becomes of similar order to the confusion noise, the noise caused by low flux density sources that are not directly detected, but sufficiently dense as to bias flux density measurements. Once this level is reached, further integration will not result in the usual decrease of noise by the inverse-square-root of the integration time. In addition to this confusion noise, there is a secondary confusion term which arises from residual sidelobes from sources that are unCLEANed because they lie below the CLEANing threshold, which in the case of CHILES Con Pol that is 5 $\mu$Jy. Additionally, any source above 5 $\mu$Jy is CLEANed down to this level, but will remain unCLEANed below this level, and thus contribute some confusing sidelobes. These sidelobes spread across the field and introduce a noise term that further increases the measured noise from thermal noise \citep{condon12}.

\begin{figure*}
\begin{center}
\includegraphics[width=0.9\textwidth]{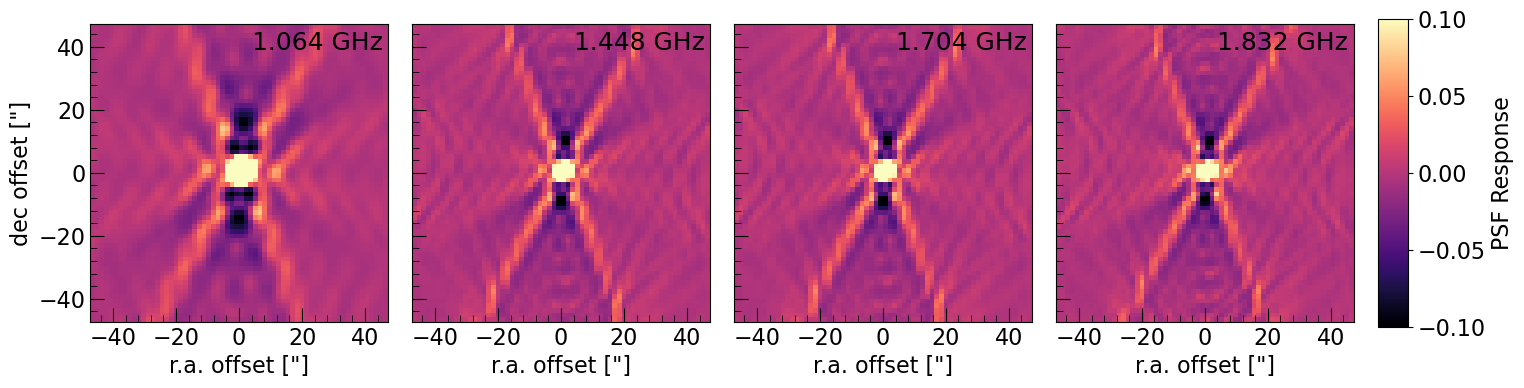}
\caption{The PSF response in each of the four subbands, whose center frequencies are indicated in the the upper right corner of each panel, with a common color scale across each panel of -0.1 to 0.1.}
\label{fig:ccp_psf}
\end{center}
\end{figure*}

\quad Interestingly, these residuals are not uniform across the field of view, but rather their prevalence decreases with increasing radial distance form the phasecenter. This is a natural consequence of the fact that there are more unCLEANed sources below the threshold in the center of the field, thus more unCLEANed sidelobes, and an increase in this confusion style-noise. The net effect of these confusion from interferometric response is to produce a, ``negative bowl," due to the strong negative features in the point-spread function of the telescope. To illustrate this, in Figure \ref{fig:ccp_psf}, we show the PSF response for each of the four subbands. Directly North and South of the peak response are regions of significant negative response, ranging from -0.09 and -0.11 for the four subbands, that create this negative bowl feature. This effect would vary with different values of robust weighting, with the worst case being for a naturally weighted image. Given the long integration of CHILES Con Pol and subsequent low thermal noise, these unCLEANed sources produce a non-negligible effect on the fidelity of our maps. Recently, \citet{mauch20} discuss this issue in their deep MeerKAT observations 1.28 GHz observations of the DEEP2 field. To correct for this negative bowl, they multiply their derived empirical primary beam pattern by the central offset which effectively flattens the image, and produces accurate flux densities. The approach we take is similar, although below, we describe simulated observations to probe specifically the role of faint unCLEANed sources.

\quad Imperative to the measurement of accurate flux densities for the CHILES Con Pol sources is the characterization and correction of this ``negative bowl" feature in our maps, which leads to a maximum systematic offset of approximately -0.4 $\mu$Jy beam$^{-1}$ at field center. In order to understand this effect, and to study its trend as a function of radial distance from the center of the field, we simulate the low flux-density regime, flux densities less than 5$\mu$Jy and greater than 0.1$\mu$Jy, and characterize the noise caused by their unCLEANed sidelobes. To do this, we first construct an image of 51.2$\arcmin$ $\times$ 51.2$\arcmin$ and populate it uniformly with point sources that have a distribution of flux densities given by the source count formalism in \citet{condon12}. This image is then primary beam attenuated, to mimic the actual response of the telescope, and all flux densities above 5$\mu$Jy, the CLEAN threshold of CHILES Con Pol, are set to 5$\mu$Jy. The map from the previous step is then used as an input model that we use to simulate the visibilities CHILES Con Pol observations using the CASA task \textit{simobserve}. We then take these visibilities and produce an image using the same parameters as with our CHILES Con Pol map, but with no CLEANing applied. This procedure provides us with a simulated image of the contribution of the confusion due to low flux density sources and sidelobes from unCLEANed sources.

\begin{figure}
\begin{center}
\includegraphics[width=\columnwidth]{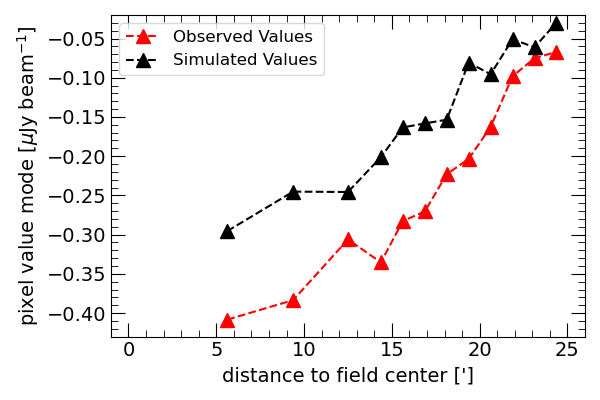}
\caption{The mean of the pixel-value distribution as a function of distance from field center for the simulated CHILES Con Pol observations of flux densities below 5$\mu$Jy (black markers), and for the CHILES Con Pol maps presented in this work (red markers).}
\label{fig:NegBowl}
\end{center}
\end{figure}

\quad Using this simulated image, we can now explore how the peak of the pixel-value distribution for the simulated map varies as a function of radial distance. To do this, we take annuli at increasing radii, taken on the center of the image, and fit a gaussian to the pixel-value distribution and save the mean of this fit as the mean pixel-value offset for this annuli. For the CHILES Con Pol image, we follow the same procedure with the same annuli in order to ensure the closest comparison. This comparison is shown in Figure \ref{fig:NegBowl}, where we show the simulated measurements (black markers) and observed measurements (red markers) as a function of radial distance. From Figure \ref{fig:NegBowl}, we see that the overall radial trend between the two is consistent, with the effect lessening at increasing radial distance. Interestingly, although the slope is consistent between the two, the values for the observed map are systematically lower. We hypothesize that this is due to the fact that the CHILES Con Pol data suffer from imperfect calibration, and thus, imperfect deconvolution. Specifically, when an image with some remaining calibration errors is being CLEANed, flux will be properly added to the CLEAN model, but the PSF structure in the image attributed to the CLEAN model flux will not be perfectly subtracted from the image, resulting in low-level artifacts. As a net result, there are more residual artifacts from these sources that are not CLEANed properly producing a larger systematic offset. Additionally, by measuring the noise in our simulated data, we estimate that the residual sidelobes are contributing approximately 50\% of the measured noise in the image.

\quad From the above investigation, we conclude that any sufficiently deep radio interferometric observation must correct for this systematic effect in a nature consistent with the observational set up of each individual observation in order to recover accurate flux densities. Especially those observations that are plagued by negative sidelobes above the 5\% level. For the CHILES Con Pol image, we correct for this offset by empirically finding the mean offset in a range of radial annuli, and adding this value to all pixels in each annuli. As a result, we produce a zero-meaned image corrected for this issue.

\section{Discussion} \label{sec:disc}

\subsection{Studying Individual micro-Jansky Radio Sources}\label{sec:microJy} 

\quad A zoom-in view of the CHILES Con Pol data shown in the top left panel of Figure~\ref{fig:ccp_fieldcompare} highlights the capability of the CHILES Con Pol data to identify individual radio continuum sources down to 5 $\mu$Jy level.  In comparison, the early science MIGHTEE 1.4 GHz survey image shown in the lower left panel has a lower resolution ($\theta=8.6\arcsec$) and a slightly lower thermal noise ($\sigma=1.7\,\mu$Jy beam$^{-1}$) while the higher resolution MIGHTEE data shown on the upper right panel has a comparable resolution ($\theta=5\arcsec$) and a significantly higher noise \citep[$\sigma=5.5\,\mu$Jy beam$^{-1}$,][]{heywood22}. Also shown for a comparison is the VLA-COSMOS 3 GHz image \citep[bottom right panel,][]{smolcic17}, which has a significantly higher angular resolution ($\theta=0.75\arcsec$) but is slightly shallower ($\sigma\approx4\,\mu$Jy at 2.3 GHz), and is at higher frequencies.

\begin{figure}
\begin{center}
\includegraphics[width=\columnwidth]{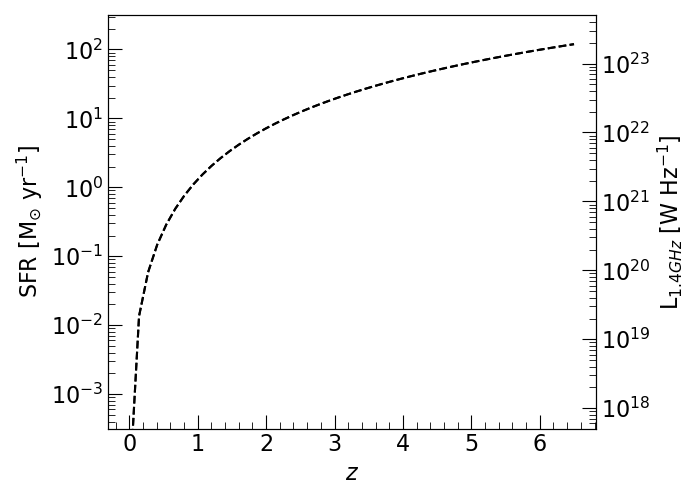}
\caption{The 5$\sigma$ detection limit, assuming a 1$\mu$Jy noise level, for the CHILES Con Pol Survey for radio continuum derived star-formation rate and the 1.4 GHz luminosity as a function of redshift.}
\label{fig:sfr_agnpower}
\end{center}
\end{figure}

\quad To translate this detection limit to galaxy properties at cosmological distances, we plot the 5$\sigma$ detection limit for star-formation rate (SFR) and the 1.4 GHz power as a function of redshift in Figure~\ref{fig:sfr_agnpower}. To calculate SFR, we use the calibration in \citet{murphy11} that relates SFR and 1.4 GHz luminosity using the FIR-Radio correlation. Our ability to identify identify and study {\em individual} star forming galaxies (SFGs) with $SFR\ge10\,M_\odot$ yr$^{-1}$ to $z=3$ is particularly important for characterizing star formation activities among the field galaxy populations in a wide range of cosmic epochs, including during the ``Cosmic Noon". These detections, and the cross-matching of them with the existing COSMOS multi-wavelength data, will allow for explorations into radio continuum science for fainter luminosities and more distant redshifts (Gim et al. in press).

\begin{deluxetable*}{lc}[t!]
\tablecaption{Summary of image noise characterization\label{tab:noise}}
\tablecolumns{2}
\tablenum{2}
\tablewidth{0pt}
\tablehead
{
\colhead{Method} &
\colhead{RMS Noise} \\
\colhead{} &
\colhead{($\mu$Jy beam$^{-1}$)} 
}
\startdata
Histogram of $R=-$1 I-image center\tablenotemark{a} & 2.04 \\
Histogram of $R=+$0.5 I-image center\tablenotemark{a} & 2.03 \\
Histogram of $R=+$0.5 I-image outside 40\arcmin radius & 1.09 \\
Histogram of $R=+$0.5 Q- and U-image & 0.79, 0.76 \\
Residual RMS noise from the dN/dS analysis & 1.67 \\
\enddata
\tablenotetext{a}{In this measurement we define image center as a square region of 8.5$\arcmin$ over the central pointing of the field, corresponding to the region shown in Figure~\ref{fig:ccp_imap_zoom}. We calculate this value on the primary beam corrected image, however, the primary beam response for our observations does not fall below 85\% allowing for us to make an estimate of image noise without significant primary beam effects.}
\end{deluxetable*}

\begin{figure*}
\begin{center}
\includegraphics[width=0.45\textwidth]{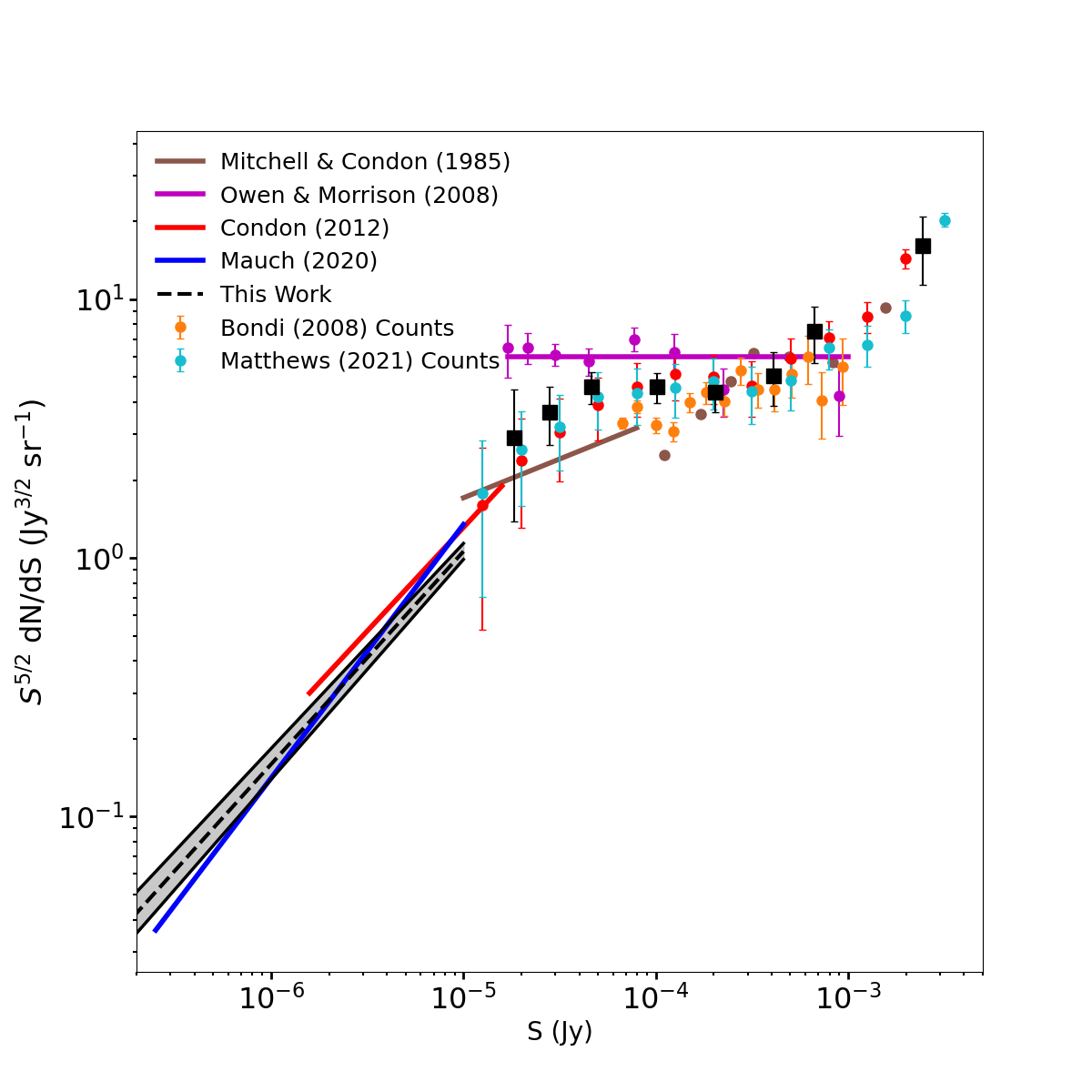}
\includegraphics[width=0.45\textwidth]{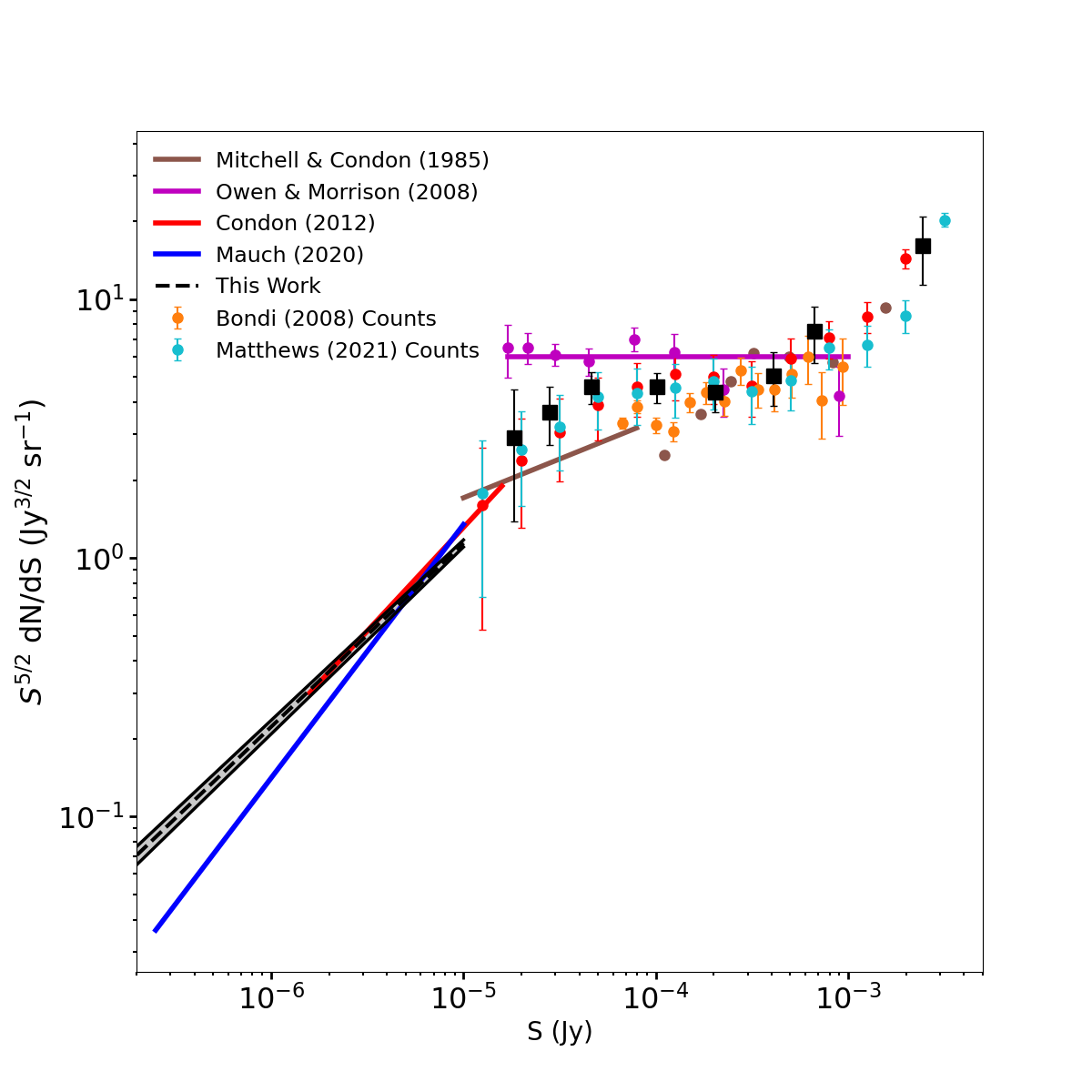}
\caption{Euclidean-normalized source counts. The 1.4~GHz $P(D)$ distribution derived from the CHILES Con Pol data for the pixel ranges of 0.1-10 $\mu$Jy beam$^{-1}$ (left panel) and 0.1-60 $\mu$Jy beam$^{-1}$ (right panel) are shown by black dashed lines. The source count derived from the CHILES Con Pol component catalog is shown by large black dots (Gim et al. in press). They are compared with the published $P(D)$ \citep{mitchell85,owen08,condon12,mauch20} and compilations of number counts by \citet{bondi08} and \citet{matthews21}.  }
\label{fig:dNdS}
\end{center}
\end{figure*}

\begin{figure*}
\begin{center}
\includegraphics[width=\textwidth]{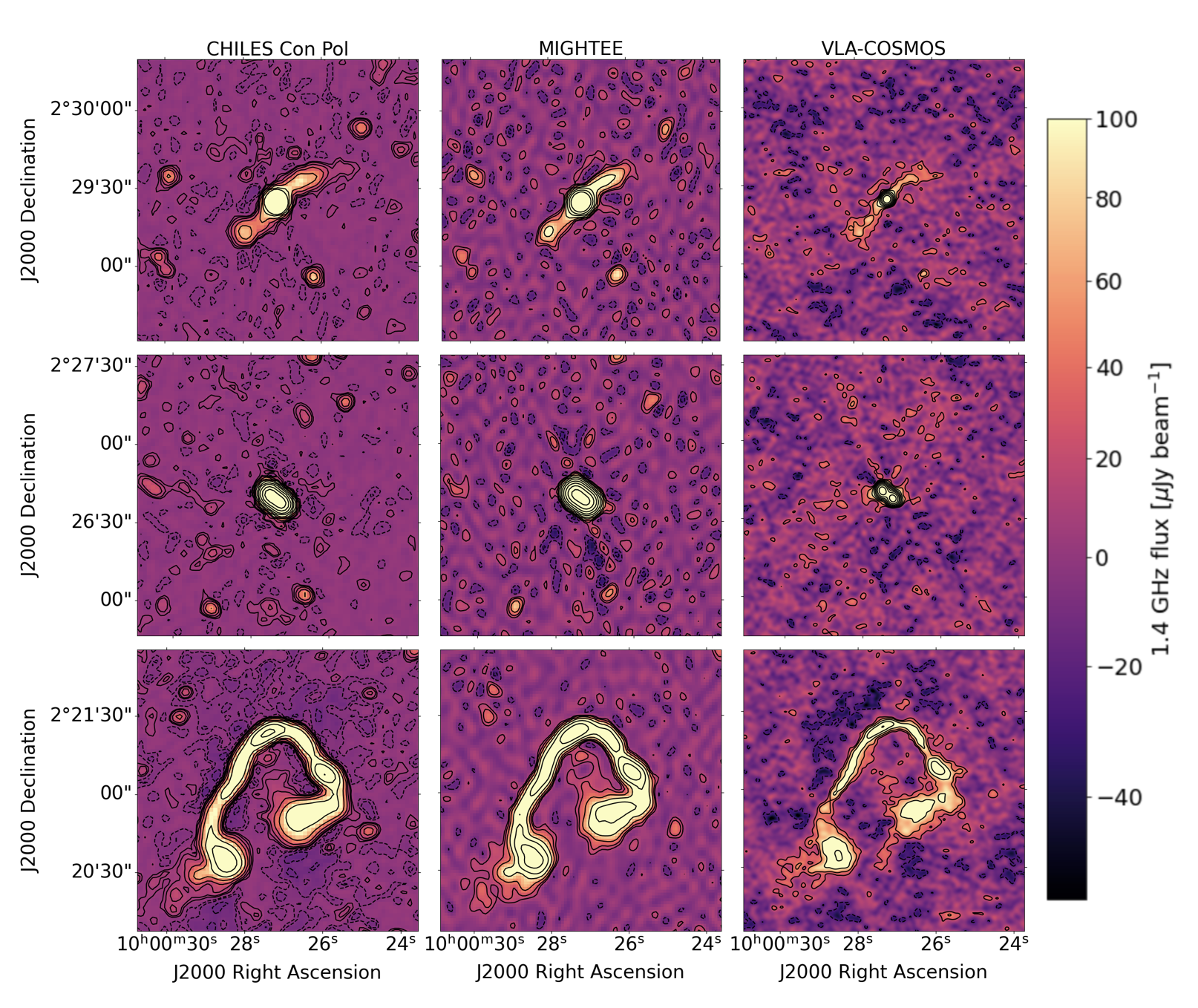}
\caption{The 1.4 GHz total intensity radio continuum images for an interesting likely AGN within the COSMOS (top) a 15 mJy source, the brightest source within CHILES Con Pol, (middle), and the ``Earmuffs" galaxy (bottom), for CHILES Con Pol, MIGHTEE \citep{heywood22}, and VLA-COSMOS \citep{schinnerer07}, from left to right. The color bar on the far right applies to all images and each map has contours at the $\pm$2$^{n}$ $\sigma$ for n values 1-11, with $\sigma$ corresponding to their respective reported rms sensitivity.}
\label{fig:ccp_compare}
\end{center}
\end{figure*}

\subsection{Statistical Inferences on the Sub-$\mu$Jy Radio Population {\label{sec:dNdS}}}

\quad Studies of source counts are a simple but powerful analysis method that can yield important insights into the cosmic evolution of an astronomical source population.  The ``$P(D)$" analysis \citep{scheuer57}, which is statistical modeling of a pixel distribution of a confusion-limited image, can provide further quantitative constraints on the source counts at flux density ranges well below the individual detection limits \citep[see a review by][and references therein]{condon12}. Following the initial characterization of the sub-mJy radio population by \citet{mitchell85} using some of the early deep radio surveys, subsequent generations of ever deeper and wider surveys have extended the radio source counts to increasingly fainter flux density levels and have characterized the nature of the radio sources as rapidly evolving populations of star forming galaxies and AGNs \citep[see][and references therein]{matthews21}. 

\quad Apparent discrepancy among different surveys around $S_{1.4GHz}\lesssim100\,\mu$Jy have raised possible systematic problems with observations and modeling assumptions, such as angular resolution and flux sensitivity, source size and clustering, sample variance, etc. \citep[see][]{bondi08,owen08,dezotti10,heywood13}.  Possible extra sources of extragalactic radio background by ARCADE 2 experiments \citep{fixsen11} also raised new interests in the faint radio source counts and their contribution to the cosmic radio background \citep{seiffert11,vernstrom11}.

\quad The Euclidean-normalized source counts derived from our $P(D)$ analysis of the CHILES Con Pol data are shown in Figure~\ref{fig:dNdS}, and they confirm the steepening of the number counts below $S_{1.4GHz} = 100\,\mu$Jy as reported by recent studies such as by \citet{condon12} and \citet{mauch20}. Our source counts derived for all pixels fainter than 10 $\mu$Jy shown on the left panel nearly exactly matches the source counts derived by \citet{condon12} using a VLA 3 GHz survey specifically designed for this purpose.  Our source counts ($dN/dS=13746 S^{-1.677}$ Jy$^{-1}$ sr$^{-1}$) is slightly steeper than the recent source counts reported by \citet{mauch20} from their 1.28 GHz MeerKAT survey ($dN/dS=107000 S^{-1.52}$ Jy$^{-1}$ sr$^{-1}$) of the DEEP2 field \footnote{Our Euclidean-normalized source counts $S^{5/2}n(>$$S)$ shown in Figure~\ref{fig:dNdS} is an integral count and is flatter in comparison.}. \citet{mauch20} analyzed the same flux density range using a least-squares power-law fit with a fixed image noise of $\sigma_n=0.55\,\mu$Jy.  These authors did not offer a detailed descriptions of uncertainties in their model parameters, but their uncertainties shown in their Figure~14 are large enough to be consistent with the \citet{condon12} results, and thus in turn also with our results.  

\quad This comparison is interesting because the CHILES Con Pol data can, in principle, probe deeper into the fainter flux levels ($S_{1.4GHz}\le 5\,\mu$Jy) by lowering the confusion limit with a higher angular resolution, as discussed in Section~\ref{sec:microJy}.  We have also taken a different mathematical formulation for the $P(D)$ analysis, using a L\'evy $\alpha$-stable distribution \citep{herranz04} and a Markov Chain Monte Carlo (MCMC) model fitting, rather than the more commonly used method by \citet{condon74}, in order to obtain more robust constraints on the modeling parameters (see the Appendix for a full description). Therefore, this steepening of the power-law slope below $S_{1.4GHz} = 100\,\mu$Jy seems to be a robust result.

\quad Source confusion depends on the actual source number counts and observed angular resolution, and the classic source count based on individual detections is presented in the catalog paper (Gim et al., in press).  We repeated the same P(D) analysis to a high flux density range, up to 60 $\mu$Jy (right panel of Figure~\ref{fig:dNdS}) and 100 $\mu$Jy (not shown but see Table~\ref{tab:MCMC}) in order to test the robustness of the standard assumption of a single power-law, which may be hampered by an inflection point between 20$\mu$Jy and 100$\mu$Jy, and a confusion limit. Both the derived power-law index and the normalization increased when the pixel ranges are expanded to higher flux density levels. In hindsight, this is not surprising since the source counts model by \citet{mitchell85} found an even steeper slope with a higher normalization when they analyzed a higher flux density range (between 10 and 86 $\mu$Jy), indicating a significant change in the shape of the source counts.  As discussed by \citet{mauch20} in their choice of a power-law model, this is probably not a robust mathematical form.  We also note that the right flux density range to model was not {\em a priori} obvious when we started this work, and this changing shape of the source counts is something future studies should keep in mind.

\vspace{1cm}

\subsection{Imaging Extended Sources{\label{sec:extended}}}

\quad To highlight the significance of the sensitivity and resolution of CHILES Con Pol, we compare our total intensity image to those presented in other surveys in the COSMOS field, specifically the Deep VLA-COSMOS Survey \citep{schinnerer07} and the MIGHTEE Survey \citep{jarvis16}. In brief, the VLA-COSMOS survey is a mosaic over the entire COSMOS field at 1.4 GHz taken with the VLA in A-configuration that achieves an rms noise of approximately 12 $\mu$Jy beam$^{-1}$ at 1.5$\arcsec$ resolution (see \citet{schinnerer07} for further details). The MIGHTEE survey is a series of four mosaics, including a field approximately the entire size of the COSMOS field, with the MeerKAT telescope at 1.4 GHz. In the COSMOS field, the MIGHTEE survey achieves an rms noise of 2.2 $\mu$Jy beam$^{-1}$ at 5$\farcs$2 resolution and 1.6 $\mu$Jy beam$^{-1}$ at 8$\farcs$6 resolution (see \citet{hale25} for more details).

\quad We measured our rms noise to be 1.3$\mu$Jy beam$^{-1}$ at 4.5$\arcsec$ resolution, when measured via a pixel histogram far from field center. This translates to a factor of approximately two better sensitivity than MIGHTEE at similar resolutions, and a factor of nine better sensitivity than VLA-COSMOS, but at a factor three worse resolution. These differences in sensitivity, survey size, and resolution result in these three surveys being complementary data sets for different scientific pursuits. CHILES Con Pol has better point source sensitivity than MIGHTEE and will therefore be able to better detect individual star-forming galaxies and AGN at higher redshifts and will be less affected by confusion noise. However, the primary beam of MeerKAT is significantly larger than the VLA's, the are covered by the 50\% point at 1.4 GHz is approximately 3.5 times larger, which allowed for MIGHTEE to probe a larger area and observe interesting objects not seen by CHILES Con Pol, such as those in \citep{delhaize21}. Both MIGHTEE and CHILES Con Pol have better sensitivity than VLA-COSMOS, but the excellent resolution of VLA-COSMOS allows for the data to be used to show smaller features, such as the complex structures in AGN jets, such as the exquisite detail in jets of the ``Earmuffs" galaxy, an example of a wide-angle head tail galaxy \citep{smolcic07}. 

\quad In order that to further compare our results to that of MIGHTEE and VLA-COSMOS, we compare the three surveys for two sources in Figure~\ref{fig:ccp_compare}. In the top row of Figure~\ref{fig:ccp_compare}, we show the CHILES Con Pol, MIGHTEE, and VLA-COSMOS, respectively from left to right, images centered on an interesting extended source present in all three images. In these, we see that all three surveys detect the extended source, but that CHILES Con Pol also has more detections of point sources around it than VLA-COSMOS, and of higher signal-to-noise than MIGHTEE. In the middle row of Figure~\ref{fig:ccp_compare}, we show images over the strongest in-field CHILES Con Pol source (approximately 15 mJy at 1.4 GHz) for the three surveys. In the bottom row, we see an example in which CHILES Con Pol has the strongest artifacts, demonstrated by the negative residuals surrounding the positive flux. However, we are still able to detect extended features at sufficient signal-to-noise compared to the other surveys, as well as recover the point sources around the source, evident in both CHILES Con Pol and MIGHTEE. The three surveys each show some residual imaging artifacts, seen as stronger positive and negative contours surrounding the source, but these artifacts are of a similar level indicating that each survey reaches similar acceptable levels of accuracy in calibration.

The imaging approach we used in imaging our field is non-traditional, as opposed to the methods used for MIGHTEE and VLA-COSMOS. More specifically, for MIGHTEE the team used second and third generation calibration techniques, self-calibration and direction dependent mitigation techniques, in order to correct for sources of error in their field \citep{heywood22}. Our utilization of uv-baseline fitting and subtraction, a technique common for spectral line data \citep{cornwell92}, of sources out of the field is significantly different. However, a comparison of images show that our technique is able to just as successfully produce images of high dynamic range, with artifacts from these sources being no higher than that of those in the MIGHTEE data. One problematic source of ours, the aforementioned ``Earmuffs" galaxy does continue to cause problems in our image, as it lies in our field, so we are unable to use this subtraction technique as we would risk subtracting flux from in-field sources. The data is, fortunately, only affected at the right ascension of the source due to the sidelobes being limited to the north-south direction, due to the fact that COSMOS is an equatorial field. 

\section{Summary} \label{sec:summary}

\quad In this paper, we introduced and described the CHILES Continuum \& Polarisation Survey. We discussed the survey design, specifically, the ability of the survey to utilize unused resources from the CHILES survey, allowing for CHILES Con Pol to obtain 1027 hours of integration time from the VLA in B-configuration. We described our calibration and imaging routines and discussed the RFI environment and mitigation techniques for our observations. Using these routines, we present the first total intensity image of the CHILES Con Pol data, which has achieved an r.m.s. noise of 1.3 $\mu$Jy per synthesized beam of size 4.5$\arcsec$ x 4.0$\arcsec$, the best sensitivity, to date, at this resolution. Our image is then discussed in the context of other 1.4 GHz surveys of the COSMOS field, where we compare and contrast the image properties of these surveys. Additionally, we compare and contrast source counts from the CHILES Con Pol data and place these results in the context of previous literature Lastly, we discussed the scientific capabilities of this dataset to probe galaxy evolution across redshift space using radio continuum data, at an unprecedented level. Future data releases will include images of all four stokes planes, and in subsequent papers we will discuss the source identification and COSMOS catalogue matching (Gim et al. in press), as well as probing the properties of these sources. \\

We thankfully acknowledge the thorough review and helpful comments from the anonymous referee that aided in improving the demonstration and presentation of results, scientific discussion, and clarity. We are very grateful for helpful discussion with, and insights from, J. H. van Gorkom.

Support for this work was partially provided by the NSF through award SOSP 21A-002 from the NRAO. The National Radio Astronomy Observatory is a facility of the National Science Foundation operated under a cooperative agreement by Associated Universities, Inc. DJP's research is partially supported by the South African Research Chair Initiative of the Department of Science \& Innovation and the National Research Foundation (NRF). DJP and NML thank the WVU Eberly College Dean's office for partial support, and partial support from NSF grant AST 1412578. MSY's research is partially supported by an NSF grant AST 1413102.

\facilities{EVLA \citep{evla}}
\software{astropy \citep{astropy}, CASA \citep{casa}}

\newpage

\appendix

\section{P(D) Analysis, Source Counts, and Source Confusion}

\quad For a confusion limited image, the deflection $D$ of pixel values from the mean $\mu$ is described as the ``probability of deflection" or $P(D)$ \citep{scheuer57}, and this quantity represents a spatial variation in the integrated total flux due to a random fluctuation in source density within an observing beam.  Assuming differential source counts of the form: 

\begin{equation}\label{dnds}
\frac{dN}{dS}=N_0S^{-\kappa},
\end{equation}

\noindent \citet{condon74} have shown that the noise-free confusion probability distribution function (PDF) is an integral and leads to a scaling relation:

\begin{equation} \label{2}
P[(N_0\Omega_e)^{1/(\kappa-1)}D]=(N_0\Omega_e)^{-1/(\kappa-1)}P(D)
\end{equation}

\noindent where the effective solid angle $\Omega_e\equiv \frac{\Omega_b}{\kappa-1}$ depends on the Gaussian beam solid angle $\Omega_b \equiv \frac{\pi\theta_1\theta_2}{4\ln2}$ ($\theta_1$ and $\theta_2$ are the full width half maximum (FWHM) widths of the elliptical Gaussian beam). The source counts within an effective beam area $\Omega_e$ is $N_0\Omega_e = \eta_1^{-1}$, where:

\begin{equation}\label{5}
\eta_1^{-1}=\frac{2\Gamma(\kappa/2)\Gamma[(\kappa+1)/2]\sin[\pi(\kappa-1)/2]}{\pi^{\kappa+1/2}}
\end{equation}

\noindent ($\Gamma$ is the factorial function). Therefore, the functional shape of P(D)  relies upon only $\kappa$ for power-law source counts.  

\quad The observed $P(D)$ is the convolution of the source confusion and image noise $\sigma_n$. Since $\Omega_e$ is derived from the normalized point source function (PSF) with a gain $G$ (i.e., $\Omega_e\equiv\int G^{\kappa-1}d\Omega$), the sidelobes of the PSF have a larger effect as $\kappa$ decreases. The rms confusion variance $\sigma_c^2$ diverges for all $\frac{dN}{dS}$, and $P(D)$ must be truncated above a cutoff deflection $D_c$ \citep{condon74,mauch20}. The confusion noise $\sigma_c$ due to sources fainter than a signal-to-noise ratio $q=\frac{D_c}{\sigma_c}$ is given by:

\begin{equation} \label{6}
\sigma_c=(\frac{q^{3-\kappa}}{3-\kappa})^{1/(\kappa-1)}(N_0\Omega_e)^{1/(\kappa-1)},
\end{equation}

\noindent and the number of beam solid angles per source brighter than $D_c$ is $\beta=\frac{q^2}{3-\kappa}$ \citep{condon12}.  This leads to the rule-of-thumb for source confusion, ``one $5\sigma$ source per 25 beam areas" for $\kappa\sim2$.

\section{Methodology} \label{sec:methodology}

\quad To accurately extract the noise-free $P(D)$ (and thus $\frac{dN}{dS}$ and $\sigma_c$), $N_0$ and $\kappa$ have to be optimized as free parameters. From the determined or parameterized $\sigma_n$, the observed $P(D)$ is extracted from the convolution of the image noise and the noise-free $P(D)$. Eq.~(\ref{2}) is ultimately derived from the $P(D)$ equation \citep{condon74}:

\begin{equation}\label{8}
P(D) = \,2\int_0^{\infty} \exp{(-N_0\Omega_e\eta_1\omega^{\kappa-1})}\cos{(N_0\Omega_e\eta_2\omega^{\kappa-1}
+2\pi\omega D)}\,d\omega
\end{equation}
where $\eta_1$ is defined in Eq.~(\ref{5}) and $\eta_2$ is
\begin{equation}\label{9}
\eta_2 = \pi^{\kappa-1/2} \Gamma[\frac{1}{2}(2-\kappa)]/\{2\Gamma[\frac{1}{2}(\kappa+1)]\}.
\end{equation}

\noindent The $D$ term in Eq.~(\ref{8}) now represents the mean deflection $\overline D$ instead of absolute zero, unless $1 < \kappa < 2$, where the sky brightness no longer diverges and Olbers' paradox can be avoided \citep{mauch20}.

\begin{figure}
\begin{center}
\includegraphics[width=0.7\textwidth]{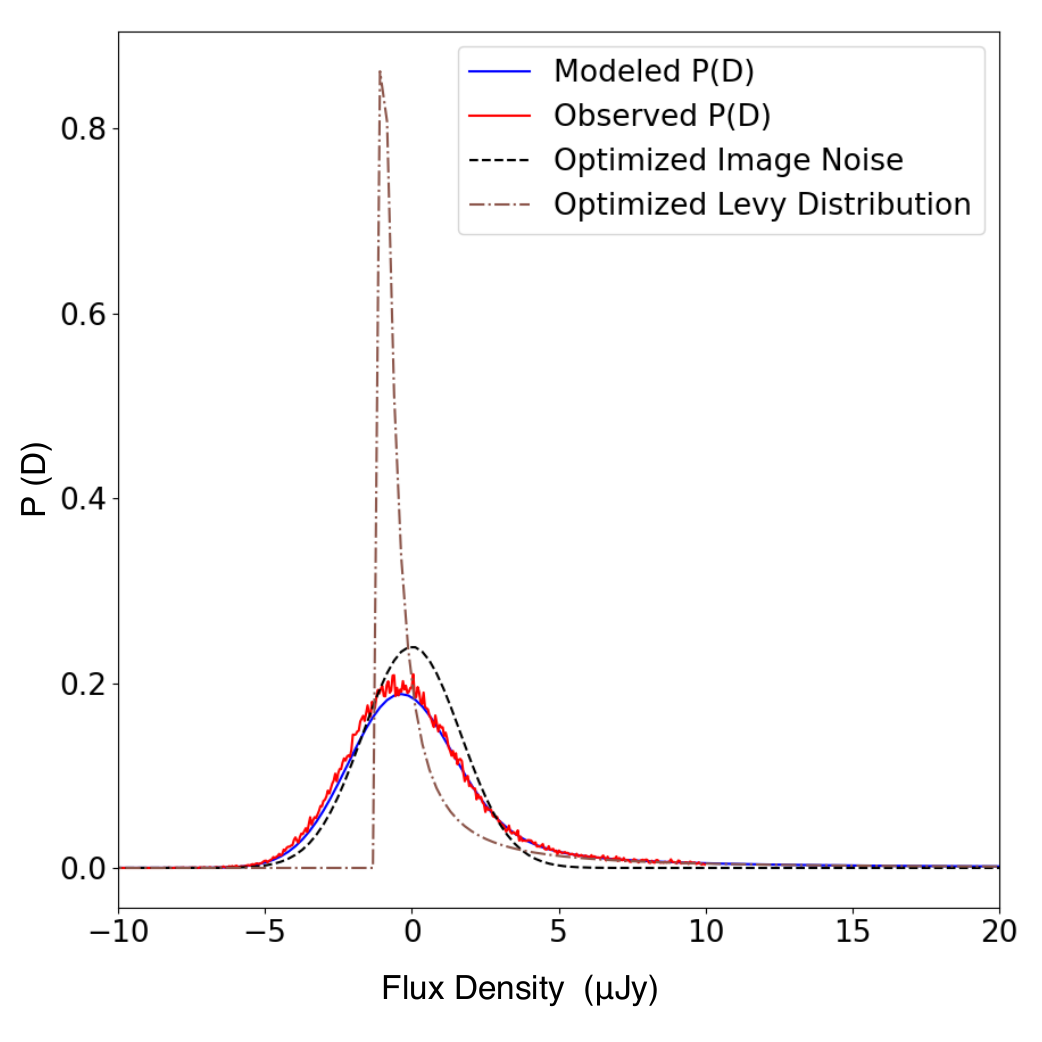}
\caption{Distribution functions of observed P(D), Gaussian noise, L\'{e}vy distribution, and modeled P(D) are shown with red solid, black dashed, black dash-dot, and blue solid lines, respectively.}
\label{fig:p_d_distribution}
\end{center}
\end{figure}

\begin{figure*}
\begin{center}
\includegraphics[width=0.45\textwidth]{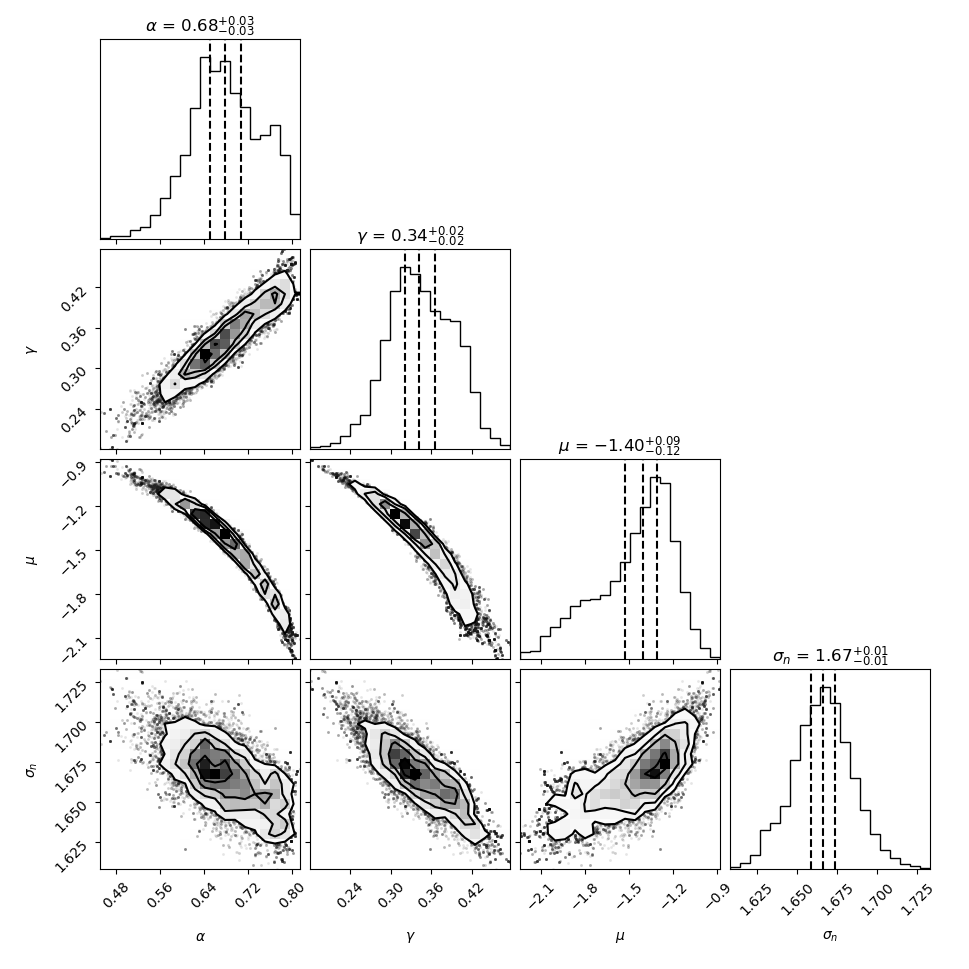}
\includegraphics[width=0.45\textwidth]{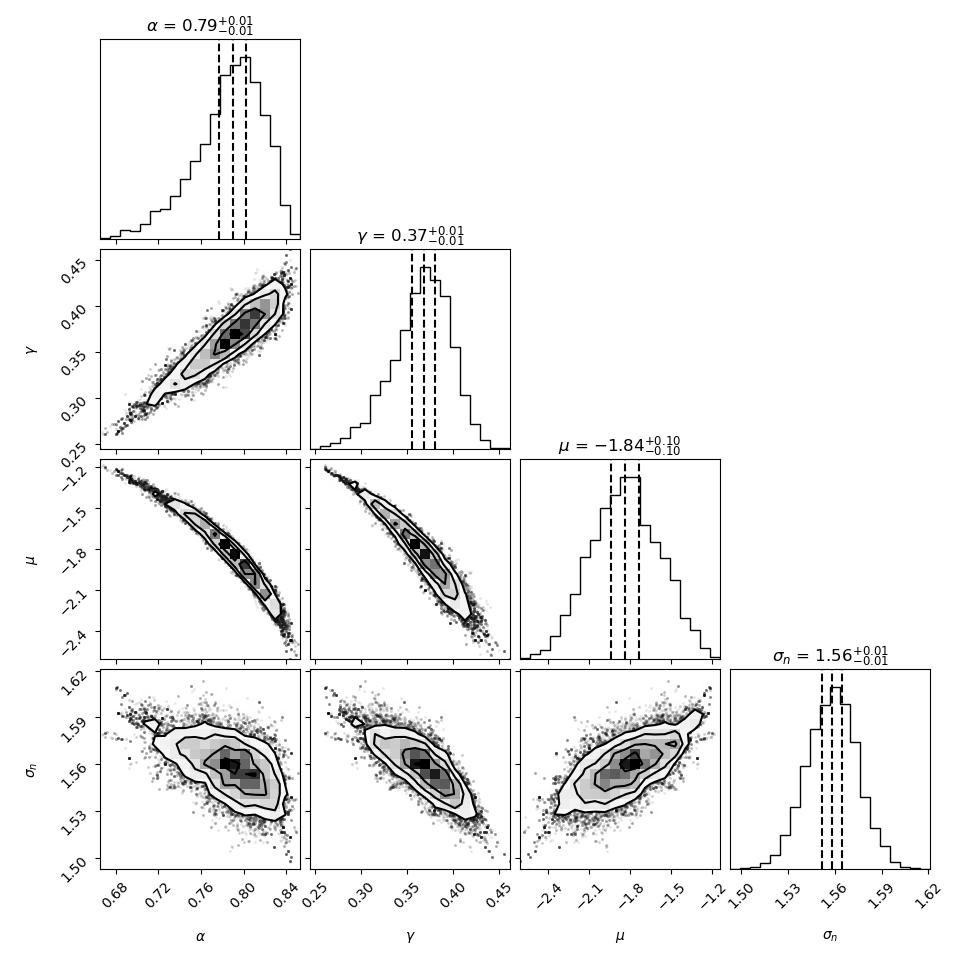}
\caption{The posterior probability distributions of the four model parameters such as $\alpha$, $\gamma$, $\mu$, and $\sigma_{0}$ are presented in 2D corner plots for models with a input pixel value range of [$-10$, $+10$] in the left and [$-60$, $+60$] $\mu$Jy beam$^{-1}$ in the right panel, respectively. }
\label{fig:cornerplot}
\end{center}
\end{figure*}

\subsection{The L\'evy Distribution}

\quad \citet{herranz04} have noted that the pixel distribution of a deep radio survey image resembles an extremely impulsive noise probability distribution function (PDF) described by a L\'evy $\alpha$-stable distribution, a statistics model popularly used in economics and engineering. The characteristic function for the L\'evy distribution completely describes any PDF, such as Gaussian or Poisson distributions, and has no analytical form. Four parameters govern the behavior of the L\'evy distribution:

\begin{enumerate}
  \item $\alpha$ is the characteristic exponent that describes the impulsiveness of the function. In the domain $\alpha = (0,2]$, $\alpha = 2$ describes a Gaussian, and as $\alpha \rightarrow 0$, the function becomes more impulsive;
  \item $\beta$ controls how skewed or symmetric the function is on the domain $\beta = [-1,1]$, where $\beta = 0$ resembles a symmetric distribution while $\beta = 1$ completely left-skews the function;
  \item $\gamma$ is the scale parameter which controls how spread out the distribution is for $\gamma > 0$;
  \item $\mu$ is the shift parameter, which reflects the the mean of the distribution.
\end{enumerate}

\noindent The L\'evy distribution is then described in the Fourier space as \citep[see][]{herranz04}:

\begin{equation}\label{10}
   \psi(w) = exp\{i \mu w - \gamma |w|^\alpha B_{w,\alpha}\}
\end{equation}
\begin{equation}\label{11}
B_{w,\alpha} = \begin{cases}
    [1 + i \beta \sgn{(w)} \tan{(\frac{\alpha \pi}{2})}] & {\rm for}\ \alpha \neq 0 \\
    [1 + i \beta \sgn{(w)} \frac{2}{\pi} \log{|w|}] & {\rm for}\ \alpha = 0.
    \end{cases}
\end{equation}

\quad A new model for $P(D)$ can then be constructed by convolving a parameter-optimized L\'evy distribution with a Gaussian distribution, representing the image noise. As a result, the L\'evy $\alpha$-stable distribution becomes a maximum-skewed ($\beta = 1$) distribution with no localization due to convolution ($\mu$ is arbitrary).

\quad We have chosen to take this formulation of $P(D)$ over the more commonly adopted form described by \citet{condon74} because this form is better suited for the Markov Chain Monte Carlo (MCMC) parameter optimization, which can yield a more robust characterization of the source counts model parameters and their uncertainties.

\subsection{P(D) Implementation}

\quad $P(D)$ is now described as the convolution of the noise-free distribution due to the underlying confused source population and Gaussian-assumed image noise. Figure~\ref{fig:p_d_distribution} illustrates these distributions as a function of flux density, depicting the observed P(D) distribution (red solid line), Gaussian noise distribution (black dashed line), and L\'{e}vy distribution (black dash-dot line). The modeled P(D) distribution (blue solid line) represents the convolution of the L\'{e}vy distribution and image noise, demonstrating excellent agreement with the observed distribution across the measured flux density range.

The parameters that define the noise-free distribution, shown in Eq.~(\ref{11}), are $\alpha$, $\beta$, $\gamma$, and $\mu$. Since $\beta$ and $\mu$ are not relevant to describing the confusion source density, only $\alpha$ and $\gamma$ are used in calculations, where

\begin{equation}\label{12}
    \alpha = \kappa - 1
\end{equation}

\begin{equation}\label{13}
    \gamma = \frac{\pi^{3/2}N_0\Omega_e}{2^{\alpha+1}\Gamma(\frac{\alpha+1}{2})\Gamma(\frac{\alpha+2}{2})\sin(\frac{\alpha\pi}{2})}.
\end{equation}

\noindent Re-arranging these equations to solve for $N_0$ and $\kappa$, the source counts parameters can be found by parameterizing and optimizing $\alpha$ and $\gamma$ as:

\begin{equation}\label{14}
    \kappa = \alpha + 1
\end{equation}
\begin{equation}\label{15}
    N_0 = \gamma\frac{2^{\alpha+1}\Gamma(\frac{\alpha+1}{2})\Gamma(\frac{\alpha+2}{2})\sin(\frac{\alpha\pi}{2})}{\pi^{3/2}\Omega_e}.
\end{equation}

Therefore, Euclidean-normalized source count is expressed as:

\begin{equation}\label{16}
S^{2.5} \frac{dN}{dS} = N_{0} S^{2.5-\kappa} = N_{0} S^{1.5-\alpha}
\end{equation}
, where $\kappa$ and $N_{0}$ are defined as Eq.~\ref{14} and Eq.~\ref{15}.

\begin{figure}
\begin{center}
\includegraphics[width=0.7\textwidth]{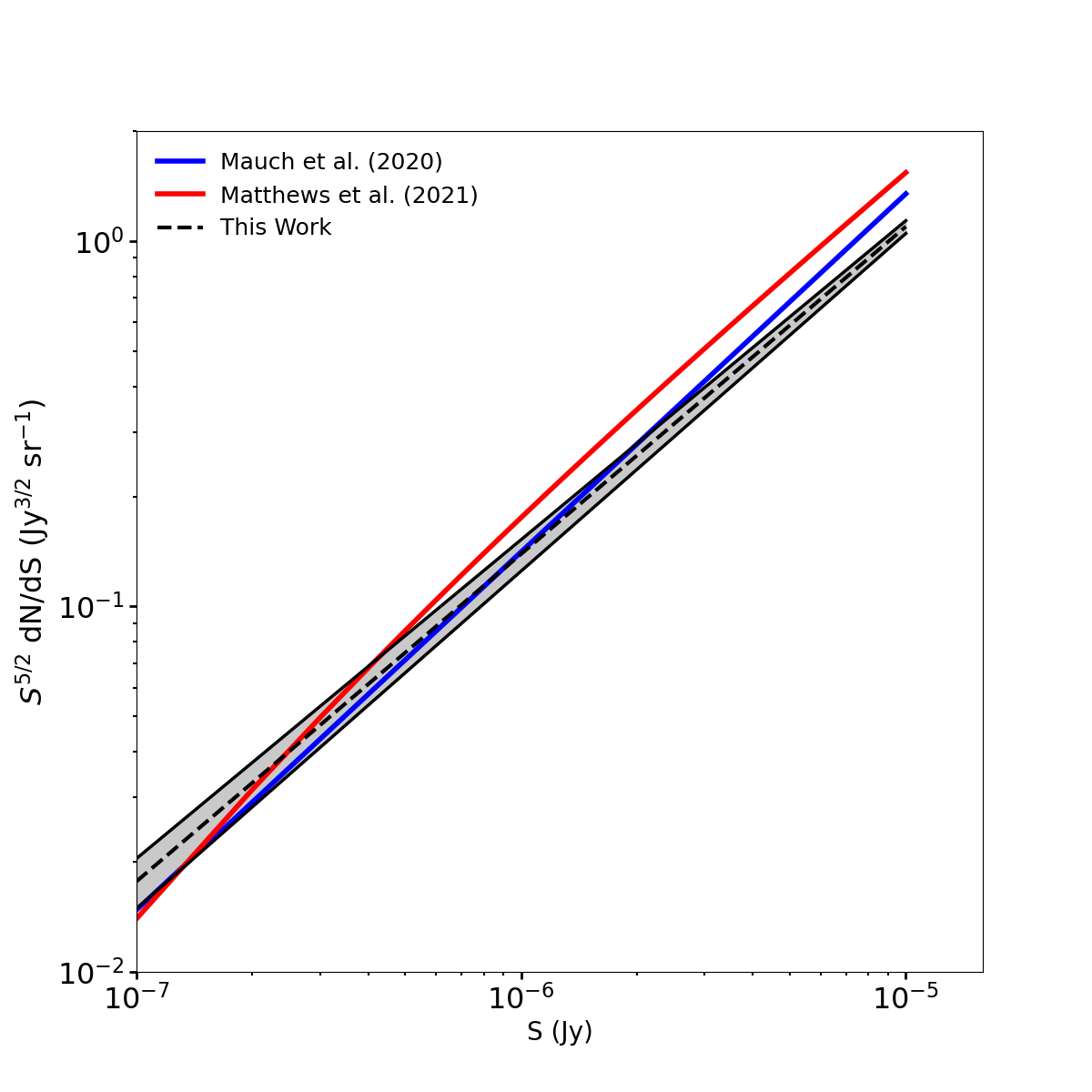}
\caption{Comparisons of Euclidean-normalized source counts with MeerKAT DEEP2 survey. Blue and red solid lines indicate the source counts obtained from \citet{mauch20} and \citet{matthews21} for the DEEP2 field observed at the MeerKAT , respectively, while the black dashed line shows the source count obtained using our P(D) model to the MeerKAT DEEP 2 image. }
\label{fig:meerkat_pd}
\end{center}
\end{figure}

\subsection{Data modeling using MCMC}

\quad The most efficient way to model $P(D)$ is to utilize the Markov Chain Monte Carlo (MCMC) method due to its effective ability to model complicated functions. The advantage of using MCMC methods over other optimization methods, such as the Nonlinear Least Squares (NLS) Regression optimizer, is that MCMC incorporates randomness to best optimize multi-parameter spaces with trial-and-error precision. The NLS Regression technique requires precise knowledge of the priors, is more susceptible to outliers, and can also get stuck on local minimums rather than the true minimum. 

\quad A downside of using an MCMC is computational intensity and time. Producing one realization of P(D) computationally takes a significant fraction of a second on a computer with a powerful CPU, which is much longer than computing the result for a much simpler function. Since optimizing $P(D)$ sufficiently requires hundreds of iterations, producing a good result can take many hours.  Therefore, a good advanced planning on the resolution of the data and the number of iterations is required. For convenience and reliability, we have utilized the python package \texttt{emcee} \citep{emcee}. This package is widely used and is well documented.  We also found it to work well with our complex function for the optimization.

\section{Results}

\quad We performed the MCMC modeling with the CHILES Con Pol image which was generated with Briggs weighting at $R=+0.5$. The primary beam attenuation was not corrected in this image. We selected pixel values within a radius of 180 pixels (corresponding to 4.5 arcminutes), ensuring minimal primary beam attenuation (6\%) while maintaining statistical robustness. 

\quad Figure~\ref{fig:cornerplot} presents the posterior probability distributions of the four model parameters ($\alpha$, $\gamma$, $\mu$, and $\sigma_{0}$) for input pixel ranges of [$-10$, $+10$] and [$-60$, $+60$] $\mu$Jy beam$^{-1}$, respectively. The corresponding best fit number count models are illustrated in Figure~\ref{fig:dNdS}, along with observational data from the literature and the CHILES Con Pol survey (Gim et al. in press). The best-fit model parameters are also summarized in Table~\ref{tab:MCMC}, including the model fitting results for the input pixel range of [$-100$, $+100$] $\mu$Jy beam$^{-1}$ (no plots are shown). 

\quad It should be noted that the best-fit model parameters are sensitive to the initial parameters, particularly for the input range of [$-10$, $+10$] $\mu$Jy beam$^{-1}$. This sensitivity may be attributed to the presence of multiple local optima. To address this, we tested various initial parameters and identified the best-fit model parameters by quantifying the difference between observed $P(D)$ and $P(D)$ constructed from the best-fit model parameters. 

\begin{deluxetable}{lcccc}
\tablecaption{Summary of the best-fit model parameters\label{tab:MCMC}}
\tablenum{A1}
\tablewidth{0pt}
\tablehead
{
\colhead{Flux density range} &
\colhead{$\alpha$} &
\colhead{$\gamma$} &
\colhead{$\mu$} &
\colhead{$\sigma_0$} 
}
\startdata
$0.1 \le S_{1.4GHz} \le 10\,\mu$Jy &
$0.677^{+0.028}_{-0.030}$ &
$0.343^{+0.021}_{-0.023}$ &
$-1.404^{+0.122}_{-0.091}$ &
$1.666^{+0.008}_{-0.007}$ \\
$0.1 \le S_{1.4GHz} \le 60\,\mu$Jy &
$0.790^{+0.013}_{-0.012}$ &
$0.369^{+0.013}_{-0.012}$ &
$-1.837^{+0.105}_{-0.099}$ &
$1.558^{+0.0007}_{-0.006}$ \\
$0.1 \le S_{1.4GHz} \le 100\,\mu$Jy &
$0.854^{+0.011}_{-0.012}$ &
$0.424^{+0.012}_{-0.013}$ &
$-2.576^{+0.214}_{-0.179}$ &
$1.538\pm0.008$ \\
\enddata
\end{deluxetable}

\quad Furthermore, we extended our P(D) analysis to the MeerKAT DEEP2 image. Following the methodology established by \citet{mauch20}, we applied the same code to the MeerKAT DEEP2 image within a 1250-pixel radius and flux density range of $-10 < S < 10$~$\mu$Jy. Comparisons with the prior studies using the same observational data demonstrate consistency between models, as illustrated in Figure~\ref{fig:meerkat_pd}, which presents Euclidean-normalized source counts from \citet[][blue solid line]{mauch20}, \citet[][red solid line]{matthews21}, and our P(D) distribution model (black dashed line). While our P(D) model exhibits a moderately shallower slope, it remains consistent with \citet{mauch20}'s model within $1\sigma$ uncertainty bounds. The model presented by \citet{matthews21} diverges slightly from both our model and that of \citet{mauch20}, particularly above 1~$\mu$Jy. This discrepancy likely stems from differing source count formulations; specifically, our approach and that of \citet{mauch20} implemented power-law models for source counts, whereas \citet{matthews21} employed a cubic polynomial function approximation.

\vspace{1mm}

\bibliography{references}{}
\bibliographystyle{aasjournal}

\end{document}